\documentclass[twocolumn]{aastex63_mod}

\usepackage{xspace}
\usepackage{amsmath}
\usepackage{hyperref}
\usepackage{breqn}
\usepackage[figuresright]{rotating}
\newcommand{\msun}{\hbox{$M_\odot$}\xspace}

\newcommand{\flux}{\hbox{$\text{erg cm}^{-2} \text{ s}^{-1}$}\xspace}
\newcommand{\lum}{\hbox{$\text{erg s}^{-1}$}\xspace}

\newcommand{\Ha}{H$\alpha$\xspace}
\newcommand{\Hb}{H$\beta$\xspace}
\newcommand{\OIII}{[\ion{O}{3}] $\lambda 5007$\xspace}
\newcommand{\OIIIsimp}{[\ion{O}{3}]\xspace}
\newcommand{\NIIsimp}{[\ion{N}{2}]\xspace}
\newcommand{\OIIIl}{[\ion{O}{3}] $\lambda 4959$\xspace}
\newcommand{\OII}{[\ion{O}{2}] $\lambda 3727$\xspace}
\newcommand{\NIIt}{[\ion{N}{2}] $\lambda \lambda 6548,6584$\xspace}
\newcommand{\NII}{[\ion{N}{2}] $\lambda 6584$\xspace}

\newcommand{\jhmag}{\hbox{$m_{J+JH+H}$}\xspace}

\newcommand{\mcsed}{\texttt{MCSED}\xspace}
\newcommand{\cloudy}{C{\footnotesize LOUDY}\xspace}
\usepackage{booktabs}
\usepackage{changepage}

\date{\today}
\shorttitle{\Ha and \OIII Luminosity Functions}
\shortauthors{Nagaraj et~al}
\graphicspath{{./}{figures/}}

\begin{document}

\title{The \Ha and \OIII Luminosity Functions of $1.2<z<1.9$ Emission-Line Galaxies from HST Grism Spectroscopy}

\correspondingauthor{Gautam Nagaraj}
\email{gxn75@psu.edu}

\author[0000-0002-0905-342X]{Gautam Nagaraj}
\affiliation{Department of Astronomy \& Astrophysics, The Pennsylvania
State University, University Park, PA 16802, USA}
\affiliation{Institute for Gravitation and the Cosmos, The Pennsylvania State University, University Park, PA 16802, USA}

\author[0000-0002-1328-0211]{Robin Ciardullo}
\affiliation{Department of Astronomy \& Astrophysics, The Pennsylvania
State University, University Park, PA 16802, USA}
\affiliation{Institute for Gravitation and the Cosmos, The Pennsylvania State University, University Park, PA 16802, USA}

\author[0000-0003-4381-5245]{William P. Bowman}
\affiliation{Astronomy Department, Yale University, New Haven, CT 06511, USA}
\affiliation{Department of Astronomy \& Astrophysics, The Pennsylvania
State University, University Park, PA 16802, USA}
\affiliation{Institute for Gravitation and the Cosmos, The Pennsylvania State University, University Park, PA 16802, USA}

\author{Alex Lawson}
\affiliation{Department of Astronomy \& Astrophysics, The Pennsylvania
State University, University Park, PA 16802, USA}

\author[0000-0001-6842-2371]{Caryl Gronwall}
\affiliation{Department of Astronomy \& Astrophysics, The Pennsylvania
State University, University Park, PA 16802, USA}
\affiliation{Institute for Gravitation and the Cosmos, The Pennsylvania State University, University Park, PA 16802, USA}

\begin{abstract}

\textit{Euclid} and the \textit{Roman Space Telescope} (\textit{Roman}) will soon use grism spectroscopy to detect millions of galaxies via their \Ha and \OIII emission. To better constrain the expected galaxy counts from these instruments, we use a vetted sample of 4,239 emission-line galaxies from the 3D-HST survey to measure the \Ha and \OIII luminosity functions between $1.16<z<1.90$; this sample is $\sim 4$ times larger than previous studies at this redshift.  We find very good agreement with previous measurements for \Ha, but for \OIIIsimp, we predict a higher number of intermediate-luminosity galaxies than previous works. We find that for both lines, the characteristic luminosity, $\mathcal{L}_*$, increases monotonically with redshift, and use the \Ha luminosity function to calculate the epoch's cosmic star formation rate density.  We find that \Ha-visible galaxies account for $\sim 81\%$ of the epoch's total star formation rate, and this value changes very little over the $1.16<z<1.56$ redshift range. Finally, we derive the surface density of galaxies as a function of limiting flux and find that previous predictions for galaxy counts for the \textit{Euclid} Wide Survey are unchanged, but there may be more \OIIIsimp galaxies in the \textit{Roman} High Latitude Survey than previously estimated.








\end{abstract}


\section{Introduction} \label{sec:intro}

Since its observational discovery by \cite{Riess1998} and \cite{Perlmutter1999}, dark energy has been at the forefront of astronomical research. While the $\Lambda$CDM paradigm has been extremely successful in predicting the properties of the cosmological microwave background and reproducing observables such as large scale structure and cosmic abundances, there are many open questions remaining, especially about the nature of dark matter and the evolution of dark energy \citep[e.g.,][and references therein]{Bull2016,Amendola2018}. In order to better constrain cosmological models, we must continue accruing more accurate and precise observational probes. 

Such efforts include the use of large-scale spectroscopic surveys to measure baryonic acoustic oscillations and redshift space distortions throughout cosmic time. The most efficient mechanisms for generating these data are multi-fiber spectroscopic surveys, such as WiggleZ \citep{Drinkwater2010,Blake2011}, the Baryon Oscillation Spectroscopic Survey \citep[BOSS;][]{Dawson2013}, the extended Baryon Oscillation Spectroscopic Survey \citep[eBOSS;][]{Dawson2016} and the Dark Energy Spectroscopic Instrument (DESI) survey \citep{DESI2016,Abareshi2022}; integrated field unit (IFU) spectroscopic surveys, such as the Hobby-Eberly Telescope Dark Energy Experiment \citep[HETDEX;][]{Gebhardt2021,Hill2021}; wide-field imaging using large, comprehensive sets of narrow-band filters \citep[e.g., J-PAS;][]{Cepa2016,Salzano2021}, and slitless (grism) spectroscopic surveys, including \textit{Euclid} \citep{Laureijs2011,Laureijs2012} and the \textit{Nancy Grace Roman Space Telescope} \citep[\textit{Roman;}][]{Green2012,Dressler2012,Spergel2015}. 

Slitless spectroscopy in particular is an extremely efficient method of obtaining spectra over an entire field with no need for the pre-selection of targets. Narrow-band surveys, such as the High-$z$ Emission Line Survey \citep[HiZELS;][]{Geach2008,Sobral2009} and the NOAO Extremely Wide Field Infrared Imager (NEWFIRM) H$\alpha$ Survey \citep{Ly2011}, achieve the same efficiency but are usually restricted to a very limited slice of redshift space, and thus require many different filters to survey large volumes. 

Normal galaxies have no strong emission lines between Ly$\alpha$ at 1216\,\AA\ and \OII, so beyond $z \sim 1$, redshift surveys are most efficiently performed at near-infrared wavelengths, with lines such \OII, \Hb, \OIII, and \Ha.  While it is possible to photometrically-select $z\gtrsim 1$ objects and then refine their redshifts with follow-up spectroscopy 
\citep[e.g.,][]{Davis2003,Steidel2004,Lilly2007,DESI2016}, emission line galaxy (ELG) surveys to probe large swaths of cosmic time are most easily performed from space. GRAPES \citep{Pirzkal2004}, which used the ACS G800L grism of the \textit{Hubble Space Telescope}, as well as WISP \citep{Atek2010} and 3D-HST \citep{Brammer2012,Momcheva2016}, which used the \textit{HST}/WFC3 G102 and G141 grisms, represent some of the first efforts to create space-based ELG samples. In this study, we use the 3D-HST sample described in \cite{Nagaraj2021a} and \cite{Nagaraj2021b}, hereafter referred to as Paper~I and Paper~II, to further explore the emission-line properties of $1.2 \lesssim z \lesssim 1.9$ galaxies.  

Two near-future missions, \textit{Euclid} \citep{Laureijs2011,Laureijs2012} and \textit{Roman} \citep{Green2012,Dressler2012,Spergel2015}, will identify millions of galaxies at redshifts $0.7\lesssim z \lesssim 2.7$ using their \Ha and \OIII emission. Given the similarity in the spectroscopic survey designs of 3D-HST, the \textit{Euclid} Deep Survey, and the \textit{Roman} High Latitude Survey (see Figure 5 in Paper~I for a visual depiction of the survey limits and observed 3D-HST fluxes), we can use 3D-HST as a pathfinder for these missions.  In particular, by evaluating the \Ha and \OIII luminosity functions and biases with respect to other galaxy samples and dark matter distributions, we can estimate how many galaxies these programs will find and how well they will be able to measure cosmological parameters.

Several efforts have been made to calculate the \Ha and \OIII luminosity functions in both the local and distant universe, and these have led to deeper understanding of galaxy evolution \citep[e.g.,][]{Gallego1995,TresseMaddox1998,Sullivan2000,JonesBlandHawthorn2001,Fujita2003,Hippelein2003,Glazebrook2004,Treyer2005,Wyder2005,Ly2007,Geach2008,Shim2009,Sobral2009,Sobral2011,Ly2011,Tadashi2011,Sobral2012,Colbert2013,Pirzkal2013,Sobral2013,Khostovan2015,Mehta2015,Sobral2015,Comparat2016,Hayashi2020,Khostovan2020}.  At $z \gtrsim 1$, most of these studies are narrow-band surveys targeted at specific redshifts; while these programs involve large numbers of sources and extremely deep exposures \citep[e.g.,][]{Khostovan2020}, their ability to investigate cosmic evolution is limited due to the small volumes covered. 

Other investigations have been limited by sample size \citep[e.g.,][]{Shim2009} or spectral resolution \citep[e.g.,][where \OIIIsimp and \Hb are a blended feature]{Pirzkal2013}.  Particularly notable is the work of \citet{Colbert2013} and the follow-up study by \cite{Mehta2015}, which used data from the \textit{HST}/WFC3 Infrared Spectroscopic Parallel (WISP) survey \citep{Atek2010} to create a sample of approximately 1,000 ELGs between $0.3<z<2.3$. 

Recently, \cite{Nagaraj2021a} carefully vetted a sample of 4350 $1.2<z<1.9$ emission-line galaxies which were originally identified on the 3D-HST grism frames by \citet{Momcheva2016}.  Here we use a subsample of 3,187 sources to determine the luminosity function and equivalent width distribution of \OIII and \Ha in this redshift range.  Since the depth and resolution of the 3D-HST data are similar to the grism surveys planned for \textit{Euclid} and \textit{Roman}, we can use our measurements to refine the predictions for these studies, and improve our measurement of the amount of intermediate-redshift star formation that is occurring in emission-line galaxies.  These data will also allow us to examine the relationship between \OIII emission and star-formation rate at an epoch intermediate between the local universe, where emission from oxygen is mostly from \OII, and the redshifts studied by \citet{Bowman2021}, where \OIII dominates.

This paper is the third in series that analyses 3D-HST sources at redshifts $1.2<z<1.9$. Paper~I showed empirical relations between stellar mass and various observational and physical properties, such as absolute magnitude in a rest-frame optical filter. Paper~II focused on the relationships among stars, gas, and dust in galaxies with available mid- and far-IR data. 

Throughout this paper, we assume a $\Lambda$CDM cosmology with $\Omega_{\Lambda} = 0.69$, $\Omega_M = 0.31$ and $H_0 = 68$~km~s$^{-1}$~Mpc$^{-1}$ \citep{Bennett2013}. All magnitudes given in the paper are in the AB magnitude system \citep{Oke1974}.

\section{Data and Selection Effects}\label{sec:data}

In this section we describe the fluxes used for the \Ha and \OIII luminosity function calculations, including the critical issue of incompleteness for low-luminosity objects.

\subsection{Data} \label{subsec:data}

As the data used to compute the \Ha and \OIII luminosity functions have been described in Papers~I and II, we give only a brief overview here. From an initial list of 9341 $1.2\lesssim z\lesssim 1.9$ candidates identified on WFC3/G141 grism frames by the 3D-HST survey \citep[GO-11600, 12177, 12328;][]{Brammer2012, Momcheva2016}, we carefully identified a clean sample of 4350 ELGs brighter than the catalog's F125W (J) + F140W (JH) + F160W (H) magnitude limit of \jhmag$=26$. Since the 625 arcmin$^2$ covered by the 3D-HST survey coincides with regions of the Cosmic Assembly Near-infrared Deep Extragalactic Legacy Survey \citep[CANDELS;][]{Grogin2011,Koekemoer2011}, the fields have a wealth of photometric observations, and these data helped fuel the results of Paper~I and Paper~II. 

Paper~I describes the process we used to identify active galactic nuclei (AGN) masquerading as normal galaxies in our sample.  First, we used X-ray matching from the deep surveys of the CANDELS fields (especially GOODS-S), to identify 72 AGN in our ELG sample. Specifically, we used a cross-correlation search radius of 1\arcsec\ to identify possible X-ray counterparts to our ELGs. Any galaxy with an X-ray luminosity greater than $10^{42}$ \lum in the $2 - 10$ keV band was considered an AGN and eliminated from the analysis. Stacking of the remaining ELGs then found X-rays levels consistent with those expected from simple star formation, suggesting that the vast majority of objects remaining in our sample are normal galaxies with no obvious AGN activity.

While X-rays are very effective for AGN identification, they fail when the gas column densities are too high ($N_H\gtrsim 5-50 \times 10^{23}~\textrm{cm}^{-2}$), which tends to correspond to highly dusty, Compton-thick systems \citep[e.g.,][and references therein]{BrandtAlexander2015}. To find such objects, we used the IRAC AGN selection criteria described by \cite{Donley2012}; the method identified 50 AGN candidates (with 11 being previously excluded via their X-ray luminosity). This whittled down our ELG sample to 4239 objects. 

\cite{Bowman2019} ran nearly the same procedure on 3D-HST galaxies at $1.9\leq z \leq 2.35$. After removing AGN, they were left with a sample of 1964 ELGs with trustworthy redshifts and clean spectra.  These data are used in \S \ref{subsec:OIIILF}, where we combine the datasets to examine the evolution of the \OIII luminosity function over the full redshift range from $1.16\leq z \leq 2.35$.

\subsection{Completeness and Selection Effects}
\label{subsec:complete}

In this work, we use the \Ha and \OIIIsimp fluxes derived from the 3D-HST grism spectra by \cite{Momcheva2016}. Due to the relatively low resolution ($R\sim 130$) of the G141 grism and the morphological broadening that occurs in extended sources, the two components of \OIIIsimp, \OIII and \OIIIl, appear as a single blended feature. However, since the ratio of \OIII to \OIIIl is fixed at 2.98:1 \citep{Storey2000}, we can simply rescale the \citet{Momcheva2016} measurements to give the de-blended fluxes for primary \OIII emission line.

A larger concern is our inability to separate \Ha from the bracketing forbidden lines of \NIIt. While corrections for \NIIsimp exist in the literature \citep[e.g., the procedure and references outlined in][]{Price2014}, gas-phase metallicities defined through strong line indicators are subject to degeneracies, inaccuracies, and other issues \citep[e.g.,][]{Kewley2008}. Furthermore, the ionization balance between the first and second excited states of oxygen changes dramatically, from predominantly O$^+$ at low redshift to O$^{++}$ at $z\sim 2$ \citep[e.g.,][and references therein]{Kewley2019}. Since N$^+$ has an ionization potential only slightly less than O$^+$, we can expect a similar trend in our data. This shift in ionization balance means that the \NIIsimp/\Ha ratio must also evolve with redshift.

Another concern associated with our understanding of the data involves the issue of dust attenuation.  The wavelength dependence of attenuation has a variety of shapes in different galaxies \citep[see][and references therein]{Shivaei2020, SalimNarayanan2020}.  An accurate modeling of this behavior is difficult, especially given its degeneracy with the other parameters of the complex stellar populations that make up a galaxy's SED\null. Even at the wavelength of \Ha (6563\,\AA), dust attenuation represents an uncertainty. Given the issues associated with \NIIsimp and dust, we do not attempt to disentangle the flux of \Ha from that of \NIIsimp; instead, we present a luminosity function for the combined lines. We discuss this further in \S \ref{subsec:SFD} where we calculate the star formation rate density between $1.2 \lesssim z \lesssim 1.9$.

An extremely important issue for any luminosity function analysis is completeness.  Following \cite{Bowman2021}, we parameterize the completeness of the \OIII and \Ha detections as a function of flux, $f$, using a modified \citet{Fleming1995} function,   
\begin{align}
    F_F(f) &= \frac{1}{2}\Bigg[ 1 + \frac{\alpha_F \log(f/f_{{\rm{50}}})}{\sqrt{1 + (\alpha_F \log(f/f_{{\rm{50}}}))^2}}\Bigg] \label{eq:fleming} \\
    \tau(f) &= 1 - e^{-f/f_{\rm 10}} \\
    F_c(f) &= [F_F(f)]^{1/\tau(f)} \label{eq:modfleming}
\end{align}
In the equations, $F_c(f)$ is the completeness, $\alpha_F$ describes how quickly the completeness drops off as a function of flux, $f_{\rm 50}$ is the flux where the recovery fraction of objects is 50\%, and $f_{\rm 10}$ is the flux at which the sample is 10\% complete. As the Fleming function is completely described by $f_{\rm 50}$ and $\alpha_F$, $f_{\rm 10}$ is not an independent parameter but a quantity directly derived from Equation~\ref{eq:fleming}. An important point to note is that for G141 grism data, the behavior of the completeness curve is virtually independent of wavelength \citep[e.g.,][]{Zeimann2014,Bowman2021}. 

There are two ways to estimate the parameters $\alpha_F$ and $f_{\rm 50}$.  The first method, which was used by \cite{Bowman2021} in their analysis of the \OIII luminosity function of $1.90 < z < 2.35$ grism-selected galaxies, is to use the data themselves and simultaneously fit the galaxy luminosity function (\S \ref{sec:lumfunc}) and each field's completeness parameters to the observed distribution of emission-line luminosities.  Such a procedure is complex, since, even if $\alpha_F$ is assumed to be the same across all five CANDELS fields, the calculation still involves at least 9 separate variables (a minimum of three for the luminosity function, one value of $f_{\rm 50}$ for each field, and $\alpha_F$).  

Alternatively, it is possible to fit the completeness curves separately from the luminosity function by assuming the intrinsic flux distribution of faint emission lines is a power law.  This approach is reasonable, given the expected nature of the faint galaxy number counts, and was the method employed by \citet{Bowman2019} in their study of the physical properties of $z\sim 2$ grism-selected galaxies.  

Our experiments show that both methods yield similar results for the galaxy luminosity function.  Therefore, to avoid any degeneracies associated with high-dimensional fits and to simplify the analysis, we chose to decouple the question of completeness from the luminosity function calculation and adopt the values of $f_{\rm 50}$ and $\alpha_F$ found by \cite{Bowman2021}.  These parameters, which were derived for $1.90 \leq z \leq 2.35$ \OIII emission on the same 3D-HST frames used here, were found using the simultaneous fitting technique described above.  Since completeness should only be a function of line flux, and not depend on the specific line being observed, the \citet{Bowman2021} values should be equally applicable to our survey.

\begin{figure}
    \centering
    \resizebox{\hsize}{!}{
    \includegraphics{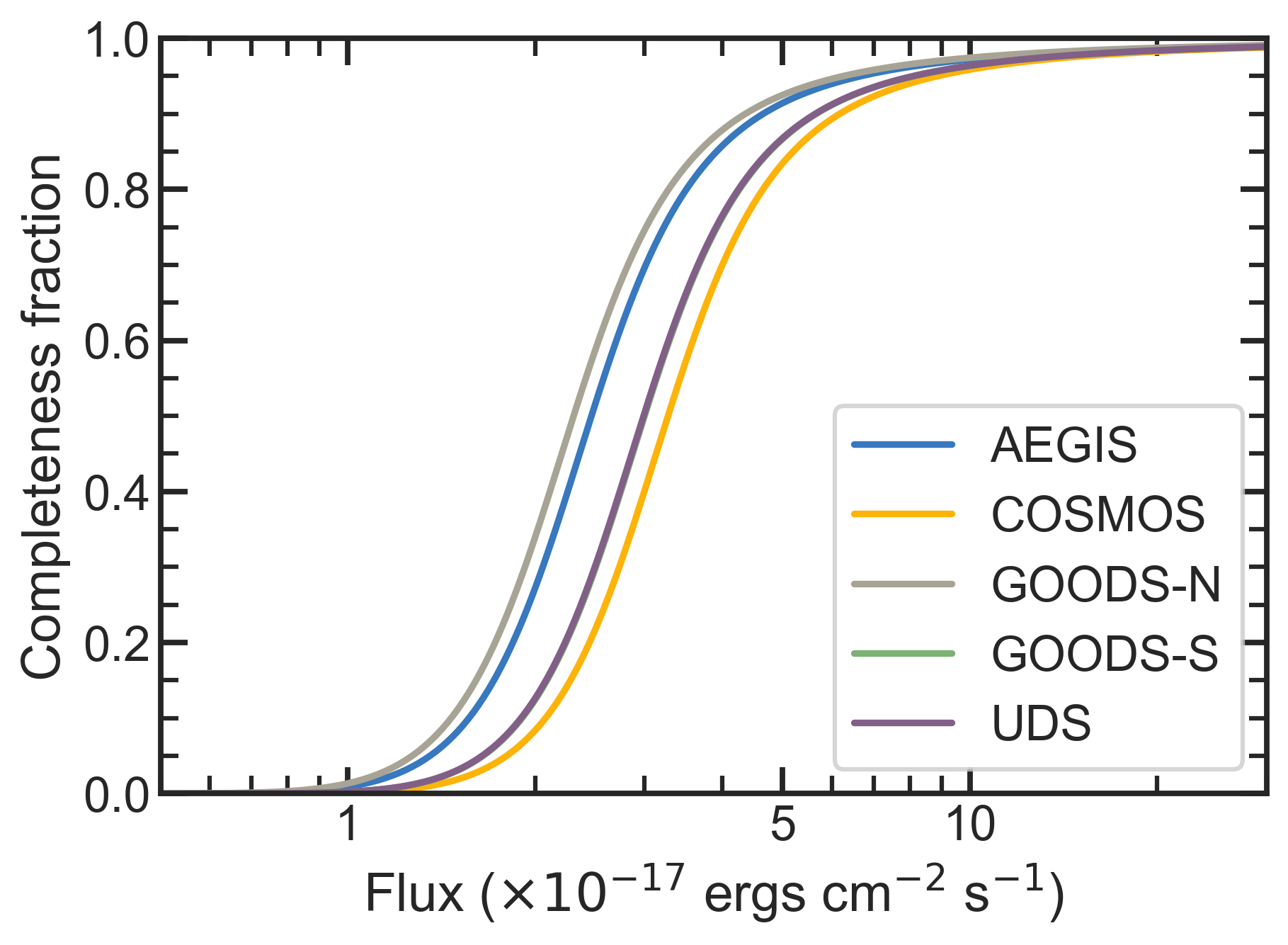}}
    \caption{Completeness curve for \Ha and \OIII in each of the five 3D-HST fields. The parameter values used to generate these curves are listed in Table \ref{tab:complete}.}
    \label{fig:HaComplete}
\end{figure}

There is one potential caveat to this last assertion.  As detailed by \cite{Momcheva2016}, the $1\sigma$ sensitivity limit of the 3D-HST survey depends not only on the strength of an emission line, but also the angular size of the emitting source. Since the present study focuses on galaxies at lower redshift than those measured by \citet{Bowman2021}, this difference has the potential to cause a shift in the survey's 50\% completeness limit.  However, based on values from the 3D-HST catalog \citep{Momcheva2016}, the mean size of $1.16 \leq z \leq 1.90$ emission-line galaxies is only 1.15 times that of the \citet{Bowman2021} systems.  In comparison, the size spread amongst the galaxies in the sample is a factor of $\sim 4$.  For this reason, we do not account for this size difference in our analysis.  We list the completeness parameters in Table \ref{tab:complete} and show the completeness curves in Figure \ref{fig:HaComplete}.


For the purposes of the luminosity functions presented in this paper, we only consider flux measurements above the 50\% levels given in Table \ref{tab:complete}.  The inclusion of objects much below this limit induces strong effects on the luminosity function by amplifying the uncertainties in the completeness curves. Conversely, if the flux limit is too strict, the sample of galaxies becomes too small for any reliable measurement of the function's low-luminosity end. Our 50\% cutoff results in 2947 \OIIIsimp and 1892 \Ha measurements being used for our luminosity function calculations. For the analysis of evolution in the \OIII luminosity function, this number can be incremented using the $1.9\leq z \leq 2.35$ ELGs found by \cite{Bowman2021}, which were identified in the same manner as the galaxies used here.  This results in a sample of 4519 \OIII ELGs above the 50\% completeness limit.

\begin{deluxetable}{lc}
\tablecaption{Completeness Parameters
\label{tab:complete} }
\tablehead{ 
\colhead{Field}
&\colhead{$f_{\rm 50}~(10^{-17}$ \flux)}
}
\startdata
AEGIS  & 2.35 \\
COSMOS & 3.12 \\
GOODS-N & 2.20 \\
GOODS-S & 2.86 \\
UDS & 2.85 \\
\enddata
\tablecomments{For all fields, $\alpha_F=4.56$.}
\end{deluxetable}


    


\section{Methodology}\label{sec:lumfunc}

\subsection{Deriving the Luminosity Function} \label{subsec:lumfunccalc}

The maximum likelihood estimator (MLE) is commonly used in astronomy to measure and fit the parameterized variables of a luminosity function. Simplifications to the MLE integral have also led to computationally efficient procedures for creating discrete luminosity functions derived from techniques such as the $1/V_{\rm max}$ \citep{Schmidt1968,Schmidt1970,HuchraSargent1973,AvniBahcall1980} and the $C^-$ \citep{LyndenBell1971} methods.  Here we do both, and derive both a digital representation of the number of galaxies versus emission-line (log) flux, and a convenient analytical representation of the luminosity function. 

We assume Poisson statistics hold, and define the likelihood of a luminosity function (scripted as $\mathcal{P}$ to avoid confusion with luminosity) following the derivation by \cite{Ciardullo2013}. For computational convenience, we use $\mathcal{L}\equiv \log L$ rather than the linear luminosity; in both cases, the math is identical, although the units are different. With the equations in this section, we are able to both measure the galaxies' discrete luminosity function and fit the data to a parameterization of our choosing
\citep[i.e., the][function]{Schechter1976}.

From \citet{Ciardullo2013}, the likelihood of observing any luminosity function, $\phi'$ is given by
\begin{equation} \label{eq:like}
    \ln \mathcal{P} = \sum_i^N \ln \phi'(\mathcal{L}_i,z_i) - \int_{z_1}^{z_2} \int_{\mathcal{L}_{\rm min}(z)}^{\infty} \phi'(\mathcal{L},z) \frac{dV}{dz} d\mathcal{L} \, dz
\end{equation}
where $N$ is the number of galaxy luminosities included in the sample.  Note that there is a  distinction between the true luminosity function $\phi(\mathcal{L},z)$, which we model as a \citet{Schechter1976} function with parameters $\alpha$, $\mathcal{L_*}$, and $\log \phi_*$, i.e.,
\begin{dmath} \label{eq:schechter}
    \phi(\mathcal{L},z) = \ln(10) 10^{\log \phi_*(z)} 10^{\left[ \mathcal{L}-\mathcal{L}_*(z)\right] \left[\alpha+1\right]} \exp \left(-10^{\mathcal{L}-\mathcal{L}_*(z)} \right)
\end{dmath}
and the observed luminosity function, $\phi'$, which is affected by incompleteness and measurement error. 

Following \cite{Marshall1983} and \cite{Marshall1985}, we define a function $\Omega(\mathcal{L},z)$ that takes into account both the flux completeness and the limits of the survey area. If $f$ is the flux, which can be thought of as a function of luminosity and redshift, and $p(f,\hat{n})$ is the completeness function in a given direction $\hat{n}$, then we have Equation \ref{eq:omega} below, where $d^2 \hat{n}$ is the integral over the unit sphere. 

In our case, we consider $p(f,\hat{n})$ constant over any given field used in the analysis. If $\Omega_i$ is the effective survey area of field $i$, then $\Omega(\mathcal{L},z)$ can be approximated as
\begin{equation} \label{eq:omega}
    \Omega(\mathcal{L},z) = \int \frac{d^2 \hat{n}}{4 \pi}\, p(f,\hat{n}) \approx \sum_{i=0}^{N_{\rm fields}} \frac{\Omega_i}{4\pi}p_i(f)
\end{equation}
Assuming that the true luminosity is isotropic, we can simply connect the observed and true luminosity functions through 
\begin{equation} \label{eq:phiphipr}
    \phi'(\mathcal{L},z) = \Omega(\mathcal{L},z) \phi(\mathcal{L},z)
\end{equation}
Going back to Equation \ref{eq:like}, $z_1$ and $z_2$ represent the survey redshift limits. The minimum luminosity, $L_{\rm min}(z)$, is somewhat arbitrary and is simply taken to be a luminosity lower than what is observable. In practice, the upper limit of the integral is also fixed at a value at, or above, that which no galaxies are expected to exist. Finally, $dV/dz$ is the differential volume element. Assuming a spatially flat universe, this is simply
\begin{equation}
    \frac{dV}{dz} = \frac{4\pi d_A^2(z)}{H(z)}
\end{equation}
where $d_A(z)$ represents the comoving angular diameter distance and $H(z)$ is the Hubble parameter.

Given this definition, the true luminosity function represents the number of galaxies per dex (log luminosity units) per comoving volume element. Moreover, given the nature of the 3D-HST survey, we assume that the flux completeness function is constant within each CANDELS field, so in a given field, $p(f,\hat{n})\equiv p(f)$. As described in \S \ref{sec:data}, we use the modified \cite{Fleming1995} completeness curve $F_c(f)$ for $p(f)$. 

The likelihood analysis presented here has not included measurement errors in the luminosity (i.e., heteroscedasticity). We follow the convention adopted in \cite{Mehta2015}, in which the luminosity errors are assumed to be normal with mean $\mathcal{L}_i$ and standard deviation $\sigma_i$. In that case, we can replace Equation \ref{eq:like} with
\begin{dmath} \label{eq:likeerr}
    \ln \mathcal{P} = \sum_i^N \ln \int_{\mathcal{L}_{\rm low}}^{\mathcal{L}_{\rm high}} \phi'(\mathcal{L}_i,z_i) N(\mathcal{L} | {\mathcal{L}_i,\sigma_i}) \frac{dV}{dz} d\mathcal{L} - \int_{z_1}^{z_2} \int_{\mathcal{L}_{\rm min}(z)}^{\infty} \phi'(\mathcal{L},z) \frac{dV}{dz} d\mathcal{L} \, dz
\end{dmath}
The drawback of this approach is that it is computationally expensive, as the normal distribution has a possibly different mean and standard deviation for every flux measurement. The results given in \S \ref{sec:LF} are therefore based on Equation \ref{eq:like}. We find that including luminosity errors has little to no effect at the bright end of the \OIIIsimp luminosity function fit but does slightly lessen the low-luminosity slope $\alpha$. In other words, the observed faint-end slope is affected by \cite{Eddington1913} bias. \cite{Bowman2021} provides a more detailed analysis of the effects of photometric uncertainties on the shape of the luminosity function.

If we assume that the luminosity function remains unchanged over the redshift interval of interest, the MLE for discrete points becomes much simpler to compute. Let us parameterize the luminosity function as a sum of discrete Dirac-delta functions at $M$ different luminosities, i.e.,
\begin{equation} \label{eq:DD}
    \phi(L) = \sum_{j=1}^{M} \phi_j \delta^D(L-L_j)
\end{equation}
Then, using Equations \ref{eq:like} and \ref{eq:DD}, the likelihood can be expressed as
\begin{equation}
    \ln \mathcal{P} = \sum_{i=1}^N \ln \phi_i - \sum_{j=1}^M \int dz \frac{dV}{dz}\Omega(L_j,z)\phi_j
\end{equation}
Setting the derivative of the likelihood to zero, the MLE solution for the luminosity function $\phi_i$ at $L_i$ becomes
\begin{equation}\label{eq:Veff}
    \phi_i^{-1} = \int dz \frac{dV}{dz}\Omega(L_i,z) \equiv V_{\rm eff}(L_i)
\end{equation}
This is a slight generalization of the historic $1/V_{\rm max}$ method \citep{Schmidt1968,Schmidt1970,HuchraSargent1973,AvniBahcall1980} for any given completeness curve $p(f,\hat{n})$. We use this formalism, which we will call the $V_{\rm eff}$ method, as a benchmark for our more sophisticated MCMC approaches.

\subsection{Computing the Luminosity Function}\label{subsec:lfcomp}

Our simplest method of computing the luminosity function is the $V_{\rm eff}$ approach introduced in \S \ref{subsec:lumfunccalc}. We calculate $\phi_i$ (Equation \ref{eq:Veff}) at every source luminosity in our sample and collect the results into luminosity (or log luminosity) bins. Wide bins allow for larger numbers of sources per bin and are thus more reliable, but do not convey as much information given their coarseness. We find that dividing the measurements into $\sim 50$ bins of equal size in log luminosity space works quite well for balancing the number of sources per bin against the complexity of the results. 

The errors on our fitted parameters are generated via a bootstrap analysis. We take the true $V_{\rm eff}$-method result using the original set of $N$($L_i, \phi_i$) values, and then generate $B$ bootstrap samples, in which the $N$ values of $L_i$ and $\phi_i$ are generated randomly with replacement. The sample variance is taken to be the error on the luminosity function measurement,
\begin{equation} \label{eq:bootstraperror}
    \hat{\sigma}^2 = \frac{1}{B-1}\sum_{i=1}^B \left[ \theta_i-\left( \frac{1}{B} \sum_{i=1}^B \theta_i  \right)  \right]^2
\end{equation}
where $\theta$ is the binned luminosity function measured at a particular interval, and $B$ is typically set to $B=100$. We include all fluxes down to the 50\% completeness limit (\S \ref{sec:data}) in the $V_{\rm eff}$ method.


For the computation of the \citet{Schechter1976} function parameters, our MCMC code uses uniform priors on $\alpha$, $\log \phi_*$, and $\mathcal{L}_*$ with bounds $[-3,1]$, $[-8,5]$, and $[40,45]$, respectively, while the completeness parameters are fixed at the values given in Table~\ref{tab:complete}. We define the relative likelihood of a solution either through Equation \ref{eq:like} (no observational errors) or Equation \ref{eq:likeerr} (with observational errors) and employ the \texttt{emcee} package \citep{ForemanMackey2013} to explore the parameter space. We compute both a static (non-evolving) luminosity function and one that evolves over time. 


To explore time evolution in the luminosity function, we use a method based on the technique described in \cite{Leja2020}. We let both $\log \phi_*$ and $\mathcal{L}_*$ be quadratic functions of redshift. However, rather than fitting the coefficients of the quadratic, whose priors are difficult to physically motivate \citep{Leja2020}, we use the values of $\log \phi_*$ and $\mathcal{L}_*$ at three specific redshifts ($z_1=1.20$, $z_2=1.76$, and $z_3=2.32$ for \OIII and $z_1=1.18$, $z_2=1.36$, and $z_3=1.54$ for \Ha) to define the quadratic formulation of $\mathcal{L}_*(z)$ and $\log \phi_*(z)$. As in our static luminosity-function calculation, the prior on $\mathcal{L}_{i*}$ is uniform over the range $[40,45]$ and the prior on $\log \phi_{i*}$ uniform on $[-8,5]$.

For the analysis, we fix the completeness parameters and also constrain $\alpha$ to be constant over time.  Doing so avoids exacerbating degeneracies between $\alpha$ and the other two Schechter parameters. (In fact, in the case of \OIIIsimp, we find that leaving $\alpha$ as a free parameter over the large redshift range $1.16 \leq z \leq 2.35$ leads to failures in the MCMC fits. For the \OIII line, we therefore we fix $\alpha=-1.5$.) 

For the reader's convenience, in Table \ref{tab:parampriors} we have listed all the parameters used in the non-evolving and redshift-varying luminosity function calculations, as well as the priors applied in the MCMC code. 

\begin{deluxetable}{ll}
\tablecaption{\Ha and \OIII Luminosity Function Fitting Parameters
\label{tab:parampriors} }
\tablehead{
\colhead{Parameter(s)} 
&\colhead{Priors} }
\startdata
$\alpha$ & Uniform on $[-3,1]$; Fixed\tablenotemark{\footnotesize a} \\
$\mathcal{L}_*$, $\mathcal{L}_{1*}$, $\mathcal{L}_{2*}$, $\mathcal{L}_{3*}$ & Uniform on $[40,45]$ \\
$\log \phi_*$, $\log \phi_{1*}$, $\log \phi_{2*}$, $\log \phi_{3*}$ & Uniform on $[-8,5]$ \\
$\alpha_F$ & Fixed \\
$f_{\rm 50}$ (5 Fields) & Fixed \\
\enddata
\tablenotetext{a}{$\alpha$ is fixed at $-1.5$ only for the redshift-varying \OIII luminosity function.}
\end{deluxetable}

In all cases, we employ 100 walkers and 1000 steps, resulting in 100,000 MCMC realizations. In other words, 100 points in the parameter space are randomly selected as initial states, and the MCMC algorithm takes 1000 steps from each initial state to find regions of higher likelihood. There is no ``best-fit'' solution, but we find that given these generous numbers for walkers and steps, the solutions generally do converge, suggesting that the true best-fit is closely approached.

\subsection{Cosmic Variance} \label{subsec:cv}

One source of uncertainty in the normalization of our emission-line luminosity functions is cosmic variance.  To estimate the expected amplitude of this effect, we use the cosmic variance calculator\footnote{ https://www.ph.unimelb.edu.au/\(\sim \)mtrenti/cvc/} of \cite{TrentiStiavelli2008}, which employs both the extended Press-Schechter formalism \citep{PressSchechter1974} and numerical simulations to compute the expected variance in any pencil-beam region of the sky. Following \cite{Colbert2013} and \cite{Bowman2021}, we compute the cosmic variance by taking the result for each of the five disconnected CANDELS fields and then dividing the average of these estimates by $\sqrt{5}$.  


The results of this calculation show that for a non-evolving luminosity function over the redshift range $1.16 \leq z \leq 1.56$, the cosmic variance expected for our \Ha-emission luminosity function  is $\sim 6.7\%$, while that for \OIII galaxies between $1.16 \leq z \leq 1.90$, this number is $\sim 5.3\%$.   For the redshift-varying case, the process of measuring the cosmic variance is not so straightforward, as we have modeled cosmic evolution using a quadratic equation, represented using the values of $\phi_*$ and $\mathcal{L}_*$ at three redshifts.  However, if we divide the surveyed redshift range into three bins, we can estimate the effect of cosmic variance on each bin.  We find that for both \Ha and \OIII, the variance should slightly increase with redshift, with the uncertainties being roughly 12\% and 9\%, respectively. 

We include the effects of cosmic variance in our calculations for the expected galaxy counts (Tables \ref{tab:HaLumFunc} and \ref{tab:OIIILumFunc}) as well as \S \ref{subsec:numcounts}) and the star formation rate density (\S \ref{subsec:SFD}).  Given the sizes of our galaxy samples and the volumes of space being surveyed ($\gtrsim 0.7$ and $1.4 \times 10^6$~Mpc$^{3}$ for the \Ha and \OIII studies, respectively), the effects of cosmic variance should be small in the case of the static luminosity function, but non-negligible for our analysis of cosmic evolution.

\section{Results} \label{sec:LF}

In this section, we present the \Ha+ \NIIsimp and \OIII luminosity functions of 3D-HST ELGs, along with the \Ha-based cosmic star formation rate density contained in the emission-line galaxies. Tables \ref{tab:HaLumFunc}, \ref{tab:OIIILumFunc}, and \ref{tab:EvolvingBest} present the overall results for our sample. 

In Tables \ref{tab:HaLumFunc} and \ref{tab:OIIILumFunc}, we list our best-fit static luminosity functions along with those of \cite{Shim2009}, \cite{Colbert2013}, \cite{Sobral2013}, \cite{Khostovan2015}, and \cite{Sobral2015}. As a cautionary statement, the luminosity functions being given in the tables are not directly comparable, as detailed in the columns labeled ``Notes''.  Moreover, because of the well-known degeneracies between the three \citet{Schechter1976} parameters, our values of $\alpha$, $\mathcal{L}$, and $\phi_i$ are not necessarily in agreement with those of the previous studies.  Nevertheless, we find that the overall form of our \Ha luminosity function is compatible with the luminosity functions derived by \cite{Colbert2013} and \cite{Sobral2013}, but somewhat distinct from the literature measurements for \OIII.  


In Table \ref{tab:EvolvingBest}, we give the parameters of our best-fit \Ha and \OIII redshift-evolving luminosity functions. For the latter, we also extend the redshift range to $z = 2.35$ using the measurements of \cite{Bowman2021}, since their galaxy sample was defined in exactly the same manner as our dataset.  


We note that in a grism survey, the true survey area is difficult to calculate as contamination from overlapping spectra and edge effects reduce the number of objects included in analyses. \cite{Bowman2021} studied this censoring by masking out those regions of the 3D-HST survey where emission-line detections are compromised, and fitting a luminosity function using only those galaxies in the unmasked areas.  They found that the effective survey area of 3D-HST is $\sim 85\%$ of the total survey area; this is consistent with the correction applied by \cite{Colbert2013} and the estimate made by \citet{Ciardullo2014}.  In this work, we reduce the quoted area of the 3D-HST survey by 15\% to approximate the effects of overlapping spectra and edge-losses.

\begin{deluxetable*}{lccccccl}
\tablewidth{0 pt}
\tabletypesize{\footnotesize}
\tablecaption{\Ha Luminosity Function Schechter Parameters
\label{tab:HaLumFunc} }
\tablehead{
\colhead{Reference} &\colhead{$z$} &\colhead{Sample Size} &\colhead{$\log \phi_*$}
&\colhead{$\mathcal{L}_*$} &\colhead{$\alpha$} &\colhead{$\log\, \int_{0.03L^*}^\infty \phi(L)\, dL$}
&\colhead{Notes} }
\startdata
\cite{Shim2009}    &0.7 - 1.9   &80   &$-2.48 \pm 0.07$        &$42.54 \pm 0.06$        &$-1.39$ (fixed)          &$-1.67 \pm 0.07$ &HST-NICMOS, \Ha \\
\cite{Colbert2013} &0.9 - 1.5   &517  &$-2.70 \pm 0.12$        &$42.18 \pm 0.10$        &$-1.43 \pm 0.17$         &$-1.8  \pm 0.2$   & WISPS, \Ha \\
\cite{Sobral2013}  &1.47        &515  &$-2.61^{+0.08}_{-0.09}$ &$42.56^{+0.06}_{-0.05}$ &$-1.62^{+0.25}_{-0.29}$  &$-1.6  \pm 0.3$   & HiZELS, \Ha, dust-corrected \\
This work          &1.16 - 1.56 &1892 &$-2.87^{+0.08}_{-0.09}$ &$42.39^{+0.05}_{-0.05}$ &$-1.60^{+0.07}_{-0.07}$  &$-1.84 \pm 0.04$\tablenotemark{\footnotesize a} &3D-HST, \Ha+\NIIsimp \\
This work          &1.16 - 1.56 &1892 &$-2.86^{+0.03}_{-0.03}$ &$42.39^{+0.02}_{-0.02}$ &$-1.60$ (fixed)          &$-1.84 \pm 0.04$\tablenotemark{\footnotesize a} &3D-HST, \Ha+\NIIsimp \\
\enddata
\tablenotetext{\footnotesize a}{Includes cosmic variance (\S \ref{subsec:cv}) in error budget.}
\end{deluxetable*}

\begin{deluxetable*}{lccccccl}
\tablewidth{0 pt}
\tabletypesize{\footnotesize}
\tablecaption{\OIII Luminosity Function Schechter Parameters
\label{tab:OIIILumFunc} }
\tablehead{
\colhead{Reference} &\colhead{$z$} &\colhead{Sample Size} &\colhead{$\log \phi_*$}
&\colhead{$\mathcal{L}_*$} &\colhead{$\alpha$} &\colhead{$\log\, \int_{0.03L^*}^\infty \phi(L)\, dL$}
&\colhead{Notes} }
\startdata
\cite{Colbert2013}   &0.7 - 1.5  &192  &$-3.28 \pm 0.09$        &$42.39 \pm 0.08$        &$-1.5$ (fixed)          &$-2.36 \pm 0.09$ &WISPS, \OIIIsimp $\lambda \lambda 4959,5007$ \\
\cite{Colbert2013}   &1.5 - 2.3  &58   &$-3.60 \pm 0.14$        &$42.83 \pm 0.11$        &$-1.5$ (fixed)          &$-2.68 \pm 0.16$ &WISPS, \OIIIsimp $\lambda \lambda 4959,5007$ \\
\cite{Khostovan2015} &1.42       &371  &$-2.61^{+0.10}_{-0.09}$ &$42.06^{+0.06}_{-0.05}$ &$-1.60$ (fixed)         &$-1.58 \pm 0.09$ &HiZELS, \Hb + \OIIIsimp \\
\cite{Sobral2015}    &1.37       &1343 &$-2.71^{+0.08}_{-0.09}$ &$42.10^{+0.05}_{-0.04}$ &$-1.60$ (fixed)         &$-1.68 \pm 0.09$ &CF-HiZELS, \Hb + \OIIIsimp \\
This work            &1.16 - 1.9 &2947 &$-2.67^{+0.06}_{-0.06}$ &$42.23^{+0.04}_{-0.04}$ &$-1.50^{+0.07}_{-0.07}$ &$-1.75 \pm 0.03$\tablenotemark{\footnotesize a} & 3D-HST, \OIII \\
This work            &1.16 - 1.9 &2947 &$-2.68^{+0.02}_{-0.02}$ &$42.24^{+0.02}_{-0.02}$ &$-1.50$ (fixed)         &$-1.76 \pm 0.03$\tablenotemark{\footnotesize a} & 3D-HST, \OIII \\
\enddata
\tablenotetext{\footnotesize a}{Includes cosmic variance (\S \ref{subsec:cv}) in error budget.}
\end{deluxetable*}

\begin{deluxetable*}{lcccccccccccc}
\tablewidth{0 pt}
\tabletypesize{\footnotesize}
\tablecaption{Redshift-Evolving Schechter Function Parameters
\label{tab:EvolvingBest} }
\tablehead{
\colhead{Line} &\colhead{$z$} &\colhead{Size} &\colhead{$\mathcal{L}_{{z_1}*}$} &\colhead{$\mathcal{L}_{{z_2}*}$} &\colhead{$\mathcal{L}_{{z_3}*}$}
&\colhead{$\log \phi_{{z_1}*}$} &\colhead{$\log \phi_{{z_2}*}$} &\colhead{$\log \phi_{{z_3}*}$}  &\colhead{$\alpha$} &\colhead{$z_1$} &\colhead{$z_2$} &\colhead{$z_3$} }
\startdata
\Ha          &1.16 - 1.56 &1892 &$42.31^{+0.09}_{-0.08}$ &$42.37^{+0.06}_{-0.06}$ &$42.52^{+0.09}_{-0.08}$ &$-2.84^{+0.11}_{-0.12}$ &$-2.81^{+0.09}_{-0.10}$ &$-3.02^{+0.10}_{-0.12}$ &$-1.60^{+0.08}_{-0.09}$ &1.18 &1.36 &1.54 \\
\OIIIsimp    &1.16 - 1.90 &2947 &$42.34^{+0.10}_{-0.09}$ &$42.30^{+0.06}_{-0.05}$ &$42.56^{+0.08}_{-0.07}$ &$-3.20^{+0.15}_{-0.18}$ &$-2.87^{+0.10}_{-0.11}$ &$-3.00^{+0.13}_{-0.15}$ &$-1.94^{+0.09}_{-0.09}$ &1.20 &1.53 &1.86 \\ 
\OIIIsimp    &1.16 - 2.35 &4519 &$42.09^{+0.04}_{-0.03}$ &$42.19^{+0.02}_{-0.02}$ &$42.88^{+0.04}_{-0.03}$ &$-2.70^{+0.04}_{-0.04}$ &$-2.47^{+0.03}_{-0.02}$ &$-3.01^{+0.04}_{-0.04}$ &$-1.50$ (fixed) &1.20 &1.76 &2.32 \\
\enddata
\end{deluxetable*}

\subsection{\Ha+ \NIIsimp Luminosity Function} \label{subsec:HaLF}

As mentioned in \S \ref{sec:intro}, the low resolution of the G141 grism and the morphological broadening associated with grism observations prevent us from separating \NIIsimp from \Ha. While prescriptions for correcting the \Ha luminosity function for \NIIsimp do exist \citep[e.g.,][]{PettiniPagel2004}, their applicability at high-redshift is uncertain.  Not only are high-$z$ metallicities generally measured via strong line indicators, which are prone to degeneracies and inconsistencies \citep[e.g.,][]{Kewley2008}, but at $z \gtrsim 1$, the ionization parameter of emission-line regions is typically larger than that seen at $z \sim 0$ \cite[see][and references therein]{Kewley2019}.  As the ionization parameter rises, the dominant form of nitrogen should shift to N$^{++}$, thereby weakening then strength of the \NIIsimp lines and decreasing their contribution to the \Ha+ \NIIsimp complex.  In this paper, we present \Ha luminosity functions that are uncorrected for the (presumably minor) contribution of \NIIsimp.

\begin{figure*}
    \centering
    \resizebox{\hsize}{!}{
\includegraphics{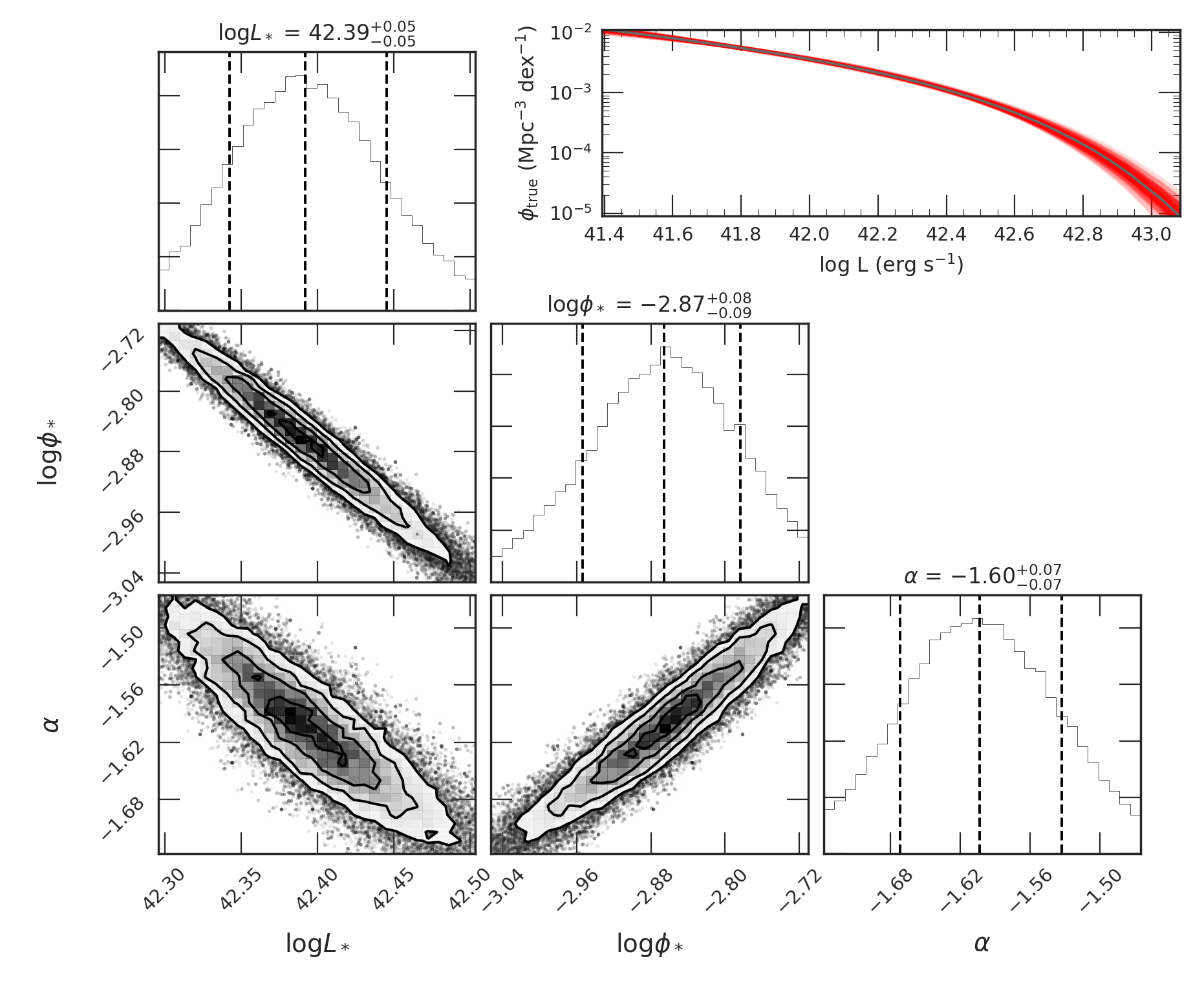}}
    \caption{Result of our Bayesian MCMC computation for the \Ha+ \NIIsimp luminosity function when assuming invariance over the $1.16 \leq z \leq 1.95$ redshift interval. The plots in the left/lower part of the figure show the marginal posterior distributions for each parameter and the 2-D cross-sections of the MCMC realizations.  As expected, all three parameters are tightly correlated; this is a common feature of Schechter fits when the data extend less then $\sim 1$~dex below $\mathcal{L}_*$.  Nevertheless, as illustrated by the plot on the top right, the shape of the luminosity function is very well-defined, with the 200 randomly chosen MCMC runs (shown in red) displaying very little scatter about the most-likely solution (plotted in gray).}
    \label{fig:HaTriangle}
\end{figure*}

In Figure \ref{fig:HaTriangle}, the top plot shows the \Ha+ \NIIsimp luminosity function with the three fitted parameters from Equation \ref{eq:schechter}: $\alpha$, $\mathcal{L}_*$, and $\log \phi_*$. Fits from two hundred MCMC iterations are shown in red, and the median fit in displayed gray. The small spread in the solutions suggests a stable and well-characterized result. 


The plots in the lower triangle show the marginal posterior distributions for the three parameters, along with 2-D contour plots of the MCMC chains. The strong correlations in the contour plots confirm that the three parameters are not independent.  This is expected, as the Schechter parameters are not orthogonal variables. However, the ubiquity of the function in the literature makes continued efforts in such a parameterization worthy.

While a non-evolving luminosity function for \Ha+ \NII is valuable, the non-negligible redshift range of the data, $1.16<z<1.56$ enables us to study the luminosity function's evolution.  As mentioned in \S \ref{subsec:lfcomp}, we did this by employing an approach based on \cite{Leja2020}, in which both $\mathcal{L}_*$ and $\log \phi_*$ are assigned to be quadratic functions of redshift.  Because of the dearth of galaxies at very low luminosities, we forced $\alpha$ to be the same across all redshifts;
this avoids the issue of degeneracies between $\alpha$, $\mathcal{L}_*$, and $\log \phi_*$ seen in Figure \ref{fig:HaTriangle}.

\begin{figure}
    \centering
    \resizebox{\hsize}{!}{
    \includegraphics{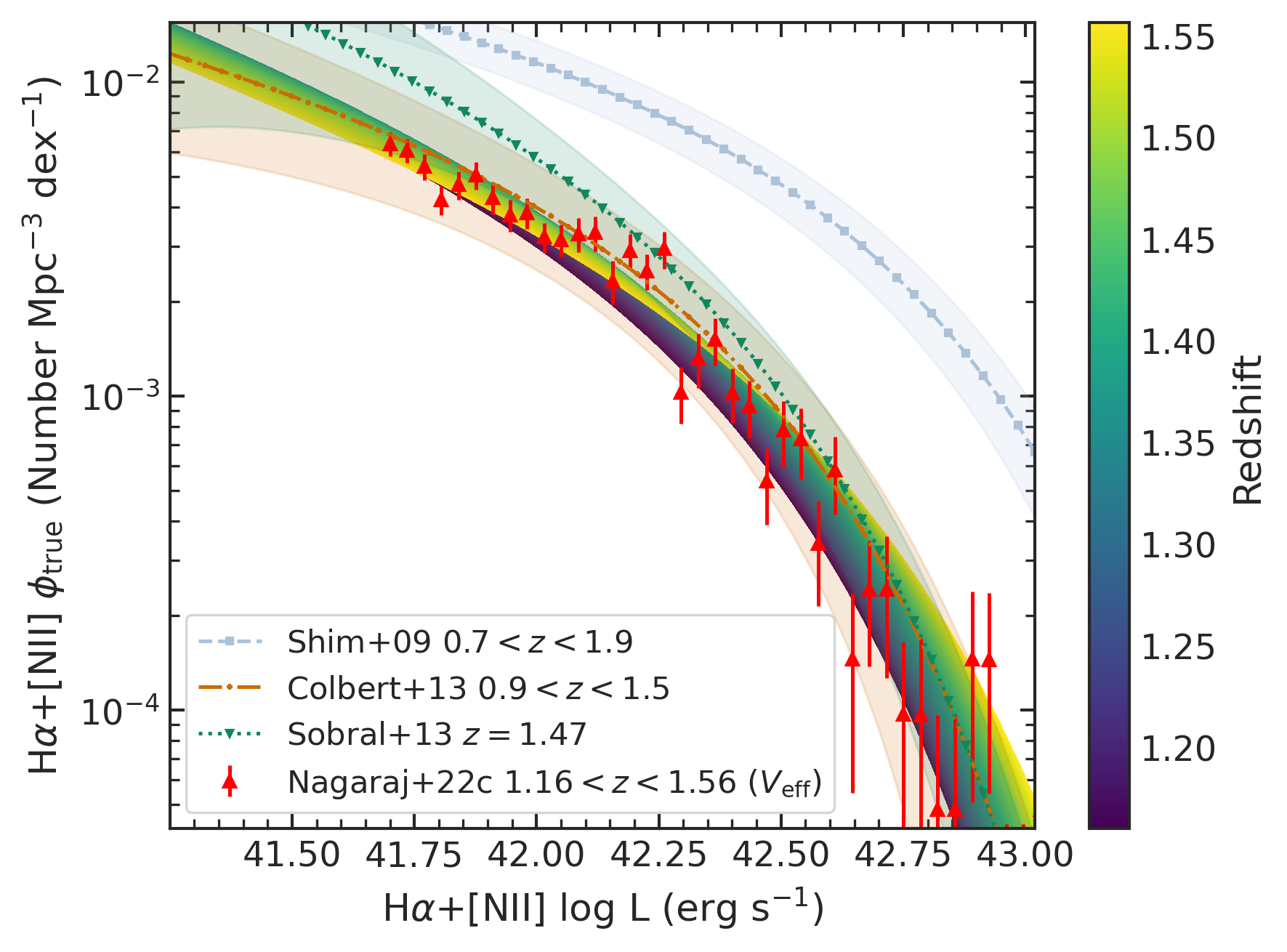}}
    \caption{Redshift evolution of the \Ha+ \NIIsimp luminosity function. The curves show the MCMC solutions for the redshift-varying \citet{Schechter1976} function, using each redshift's median parameter values (closely related to the highest-likelihood solution).  Also shown are the best-fit curves from \cite{Shim2009}, \citet{Colbert2013}, and \citet{Sobral2013}. Our results are in general agreement with the literature, especially the results of \cite{Colbert2013}. The MCMC results are also compatible with our $V_{\rm eff}$ data point (red triangles). In regard to evolution, we find that the knee of the luminosity function, $\mathcal{L}_*$, increases with redshift while the normalization factor $\log \phi_*$ decreases, although this may not be due to true redshift evolution (see text for more details). In any case, the evolution is not particularly strong in our $[1.16,1.56]$ redshift range.}
    \label{fig:HazLF}
\end{figure}

We find that the redshift-evolving luminosity function solution is quite similar to the static case, with a statistically indistinguishable low-luminosity slope $\alpha$ and similar $\mathcal{L}_*$ and $\log \phi_*$ values. All relevant quantities are presented in Table~\ref{tab:EvolvingBest}.  Figure \ref{fig:HazLF} shows the best-fit luminosity functions with redshift indicated by the color. The use of quadratic functions for $\mathcal{L}_*$ and $\log \phi_*$ allows for smooth, differentiable evolution of the luminosity function. The red points on top of the color curve show the $V_{\rm eff}$ result for the entire redshift range. The points are in greater agreement with the redshift-constant luminosity function but are still within the bounds of the redshift-varying function showed here.

We also show how our results compare to the literature. We include comparisons to \cite{Shim2009}, who found 80 \Ha emitters at redshifts $0.7<z<1.9$ through a \textit{Hubble}-NICMOS grism survey; \cite{Colbert2013}, who analyzed a sample of 517 \Ha emitters at $0.9<z<1.5$ found through the \textit{Hubble} WISP program; and \cite{Sobral2013}, who obtained a dataset of 515 \Ha emitters at $z=1.47$ through the HiZELS narrow-band imaging.  To ensure an apples-to-apples comparison, the literature luminosity functions have been modified to reflect the inclusion of \NIIsimp in our data.   \cite{Shim2009} and \cite{Colbert2013} assume $F_{{\rm H}\alpha} = 0.71F_{{\rm H}\alpha + {\rm [NII]}}$, while \cite{Sobral2013} uses a formula for correcting \NIIsimp based on equivalent width, which gives an average correction of 25\% over all their data.  We undo these corrections, along with the 1~mag internal extinction correction applied by \cite{Sobral2013}. 

Figure \ref{fig:HazLF} shows the results.  Our sample of \Ha-emitting galaxies is $\sim 4$ times larger than that of any previous study.  But from 
the figure, it is clear that the \Ha luminosity functions of \citet{Colbert2013} and \citet{Sobral2013} are in good agreement with our work.  The only serious discrepancy is with the curve produced by \citet{Shim2009}, but since that measurement was based on only 80 objects, this difference is not a concern.


From both Table~\ref{tab:EvolvingBest} and Figure \ref{fig:HazLF}, we see that $\mathcal{L}_*$ increases with redshift. Given that we are highly complete at all redshifts above $\mathcal{L} > 41.8$, this finding must reflect a physical difference as we go back in cosmic time: there are more high-luminosity ELGs at earlier epochs. From Figure \ref{fig:HazLF}, we also find fewer low-luminosity objects at higher redshifts, but this result is more subject to completeness issues, and our result may not be robust.

\begin{figure*}
    \centering
    \resizebox{\hsize}{!}{
    \includegraphics{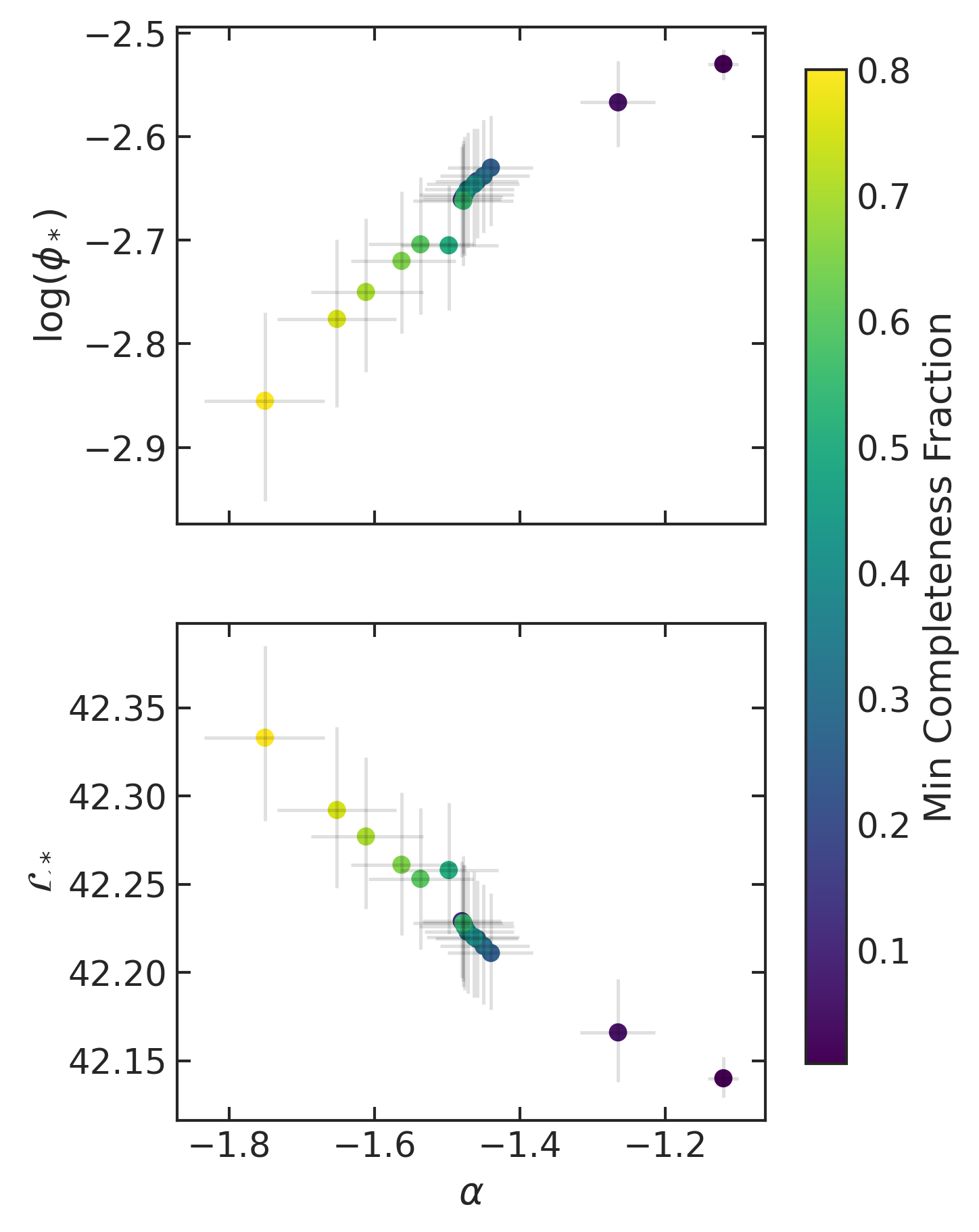}
    \includegraphics{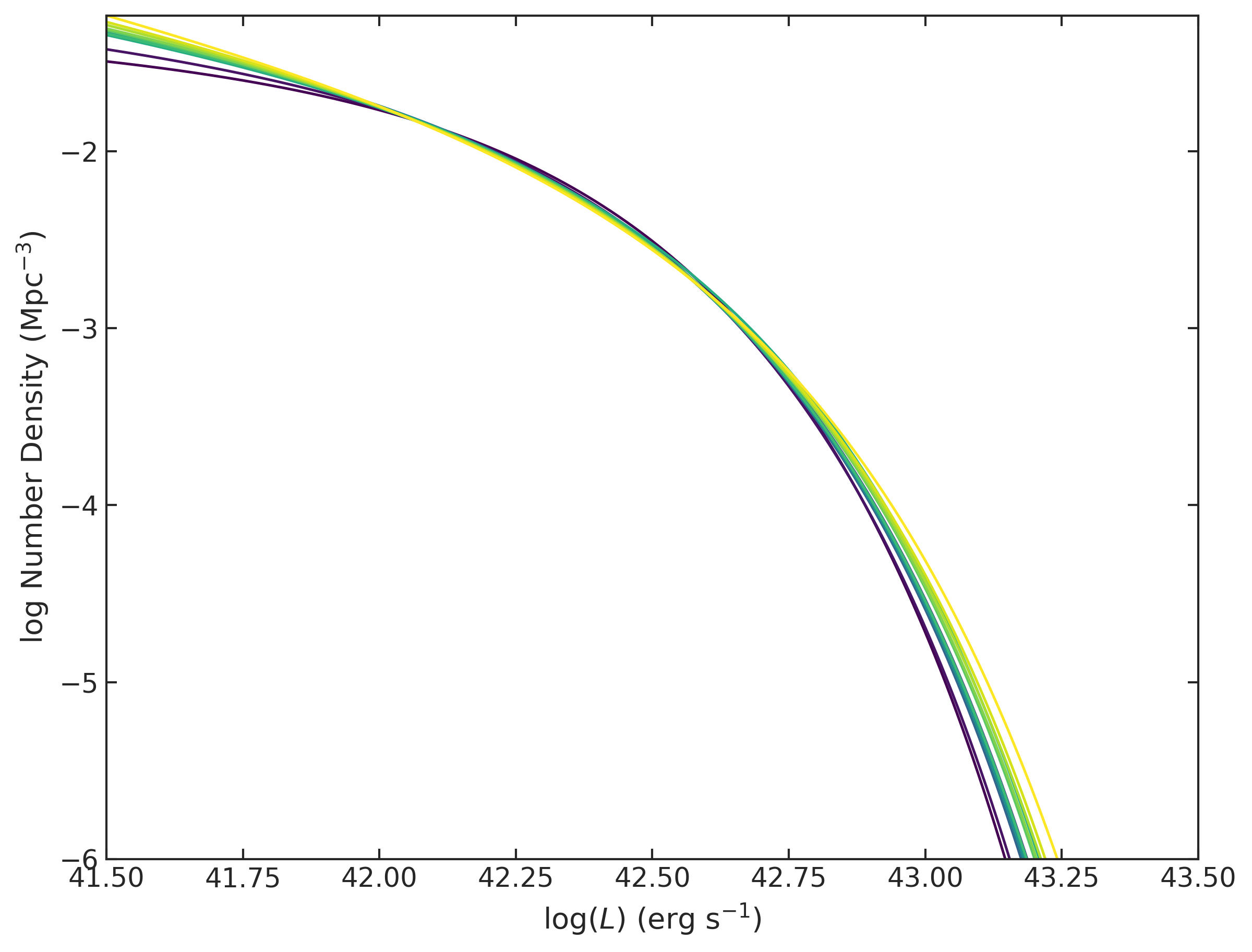}}
    \caption{Effects of the minimum completeness fraction considered for \Ha fluxes on the fitted parameters.  The left panels show that $\alpha$ decreases (gets steeper), $\mathcal{L}_*$ increases, and $\phi_*$ decreases as the completeness limit for the analysis increases. This experiment demonstrates the importance of completeness in determining the \citet{Schechter1976} function parameters. The right panel shows the luminosity functions for the Schechter parameters in the left panels. The lines are colored according to the minimum completeness fraction. We can see that the effects of completeness on the shape of the luminosity function are much less pronounced than that variables used in the parameterization.}
    \label{fig:mcf_PS_Ha}
\end{figure*}

We explore the effects of incompleteness in Figure~\ref{fig:mcf_PS_Ha}. For our main results, we have fixed the minimum completeness fraction at 50\%, i.e., we have excluded from the analysis all objects with monochromatic fluxes fainter than the 50\% limit shown in Figure~\ref{fig:HaComplete}.  But in Figure~\ref{fig:mcf_PS_Ha}, we perform an experiment in which we vary the minimum threshold from 1\% to 80\% and observe the effects on the best-fit luminosity function. As shown in the left panels of the figure, the best-fit value for $\alpha$ decreases (gets steeper), $\mathcal{L}_*$ increases, and $\phi_*$ decreases when the minimum completeness fraction increases. 

Nevertheless, from the right panel, we see that the overall luminosity function does not vary significantly in the luminosity range we are able to observe. This is because the three Schechter parameters are correlated: various sets of values can lead to the same overall luminosity function. We see from this experiment that the decision of which flux measurements to include changes the parameterization of the luminosity function, but not the overall shape curve.

\subsection{Cosmic Star Formation Rate Density} \label{subsec:SFD}

The evolution of the star formation rate density (SFRD) of the universe is an important indicator of galaxy growth and evolution. As reviewed by \cite{MadauDickinson2014}, we have the general picture that star formation peaked around $z\sim 2$, an era dubbed as ``cosmic noon,'' and has been declining at $\sim 0.1$~dex per Gyr ever since.  However, while this outline is known, further measurements of the SFRD, especially for certain populations of galaxies such as ELGs, are useful for improving our knowledge of the evolution of star formation and quantifying how selection effects propagate into this understanding. In this section, we use the \Ha luminosity function to calculate the SFRD between $1.16 \leq z \leq 1.56$.

\subsubsection{Conversion between \Ha luminosity and Star Formation Rate} \label{subsubsec:conversion}

\Ha luminosity is often used as a direct proxy for very recent star formation \citep[under 10 Myr; e.g.,][]{Kennicutt2012}. As such, applying a conversion formula from \Ha luminosity to star formation rate (SFR) is a common process. Typically, the conversion from \cite{Hao2011} and \citet{Murphy2011} as compiled by \cite{Kennicutt2012} is used. 

Nevertheless, the aforementioned relation is calibrated in the local universe, therefore raising the concern of its application to higher redshifts. For example, at lower metallicities, main sequence stars tend to be bluer and hotter, changing the amount of ionizing radiation emitted by massive stars, and thus the calibration. 

Furthermore, the \Ha-SFR calibration summarized by \citet{Kennicutt2012} applies to dust-corrected \Ha brightness, and at higher redshifts, the details of dust attenuation constitute a major source of uncertainty \citep[e.g.,][]{Bouwens2012,Nagaraj2021b}.  This error propagates directly into the SFR conversion, and is then compounded by the fact that our \Ha data is contaminated by \NIIt, while the SFR calibration of \citet{Kennicutt2012} is for \Ha only.  Thus, the applicability of the local conversion of \Ha luminosity to SFR is unclear.

We can address this issue directly by calculating our own relation between SFR and observed \Ha+ \NIIsimp luminosity. In Paper~I, we discussed the SED fitting procedure of the entire sample of $1.16 \leq z \leq 1.90$ 3D-HST emission-line galaxies. We used the Bayesian MCMC SED code \mcsed \citep{Bowman2020} to estimate the physical properties of our ELG sample, assuming a \cite{Kroupa01} initial mass function (IMF), a single-valued (but free) stellar metallicity, a binned star formation history, with the SFR in each bin a free parameter, and a dust attenuation parameterization from \cite{Noll2009} and \cite{Kriek2013}. 

\mcsed uses simple stellar population (SSP) SEDs from the Flexible Stellar Population Synthesis (FSPS) library \citep{Conroy2009,Conroy2010SPSM} with Padova isochrones \citep{Bertelli1994,Girardi2000,Marigo2008}. Nebular emission is treated via interpolation in tables of \cloudy models \citep{Ferland1998,Ferland2013} computed by \cite{Byler2017}. Gas-phase metallicity ($Z_{\rm gas}$) is set equal to the stellar metallicity in \mcsed, and we fixed the ionization parameter at $\log U = -2.5$ based on the high \OIIIsimp/H$\beta$ ratios observed for our sources.

Our SED fits are based primarily on the 3D-HST photometry collected by \cite{Skelton2014}, which consists of 147 filter bands distributed over the five CANDELS fields and covering the wavelength range from 3,000~\AA\ to 80,000~\AA\ (observed frame). In Paper~I, we merged these data with photometry from \textit{Swift} and \textit{GALEX}, which extended the wavelength coverage for over 400 sources to $\sim 2,000$ \AA\ (observed-frame). Also, mid- and far-IR measurements from \textit{Spitzer} and \textit{Herschel} were added for over 600 sources in Paper~II, but to maintain consistency within the sample, we do not include these dust-sensitive wavelengths here. In addition, \mcsed is able to employ emission line fluxes in its SED-solution, making it an ideal analysis tool  for grism-based surveys. We included \Ha, \Hb, and \OIII fluxes for galaxies whenever available.

Our \mcsed-based SFR estimates are fairly robust.  We find that the mean uncertainty on our SFRs is $0.24$ dex, with a standard deviation of $0.08$ dex. Moreover, in Paper~I, we examined how the basic assumptions underlying our SED fits affected the derived properties of the 3D-HST galaxies.  For example, we found that changing the ionization parameter, modifying the weights assigned to the emission line fluxes, or fixing galaxy metallicity all led to statistically indistinguishable distributions for the galaxies' SFRs.  In other words, ionization parameter, metallicity, and the inclusion of emission line fluxes do not noticeably affect the systematics of an SFR measurement. In addition, \cite{Bowman2020} showed that for a large sample of 3D-HST galaxies, the choice of dust attenuation curve does not strongly affect the SFR estimate.

On the other hand, we do find that the details of a galaxy's assumed star formation history (SFH) do affect our SFR estimates.  Our fits are based on a ``non-parametric'' SFH, i.e., one in which the SFR of each epoch in a galaxy's history is fit independently of the other epochs.  Such fits have been proven to reduce a bias in SFR measurements that is introduced by the use of parameterized SFHs  \citep[e.g.,][]{Conroy2013,Leja2017,Leja2019a,Bowman2020}, though the magnitude of this bias reduction depends on the prior used in each age bin \citep{Leja2019a}.  Our quoted SFRs represent the star formation rate during the most recent age bin, i.e., over the last 100 Myr of cosmic time.

Given that we are trying to calibrate the SFR-\Ha relation, the lack of influence of the \Ha flux on the SFR measurement suggests that correlations found between SFR and \Ha are not artificially induced by the method of measuring SFR\null. Furthermore, the benefit of using \mcsed SFRs is that we have a way of connecting the observed \Ha + \NIIsimp fluxes with SFRs that takes into account dust attenuation and contamination by \NIIsimp. This reduces the bias and uncertainty introduced by applying single values for the dust and \NIIsimp corrections.


To calculate the mean SFR-\Ha relation, we use a procedure very similar to that employed by \cite{Bowman2021} in their analysis of the \OIII luminosity function. We first divide the \Ha+ \NIIsimp luminosities into 25 bins with equal numbers of objects each interval. In each bin, we adopt the mean linear SFR and linear \Ha luminosity as representative values. We then take the logarithm of those values and fit a line. We chose this process because if we average the measurements in logarithmic space, we would underestimate the total SFR of the population. As suggested by \cite{FeigelsonBabu1992}, we use the orthogonal distance regression technique to fit the line, since there is non-negligible heteroscedasticity associated with the measurement errors for both luminosity and SFR.

To estimate errors in our solution, we bootstrap the \Ha+ \NIIsimp luminosity measurements and repeat the procedure above 1000 times. This error estimation process is the same as that described in \S \ref{subsec:lfcomp} for the $V_{\rm eff}$ method.

\begin{figure}
    \centering
    \resizebox{\hsize}{!}{
    \includegraphics{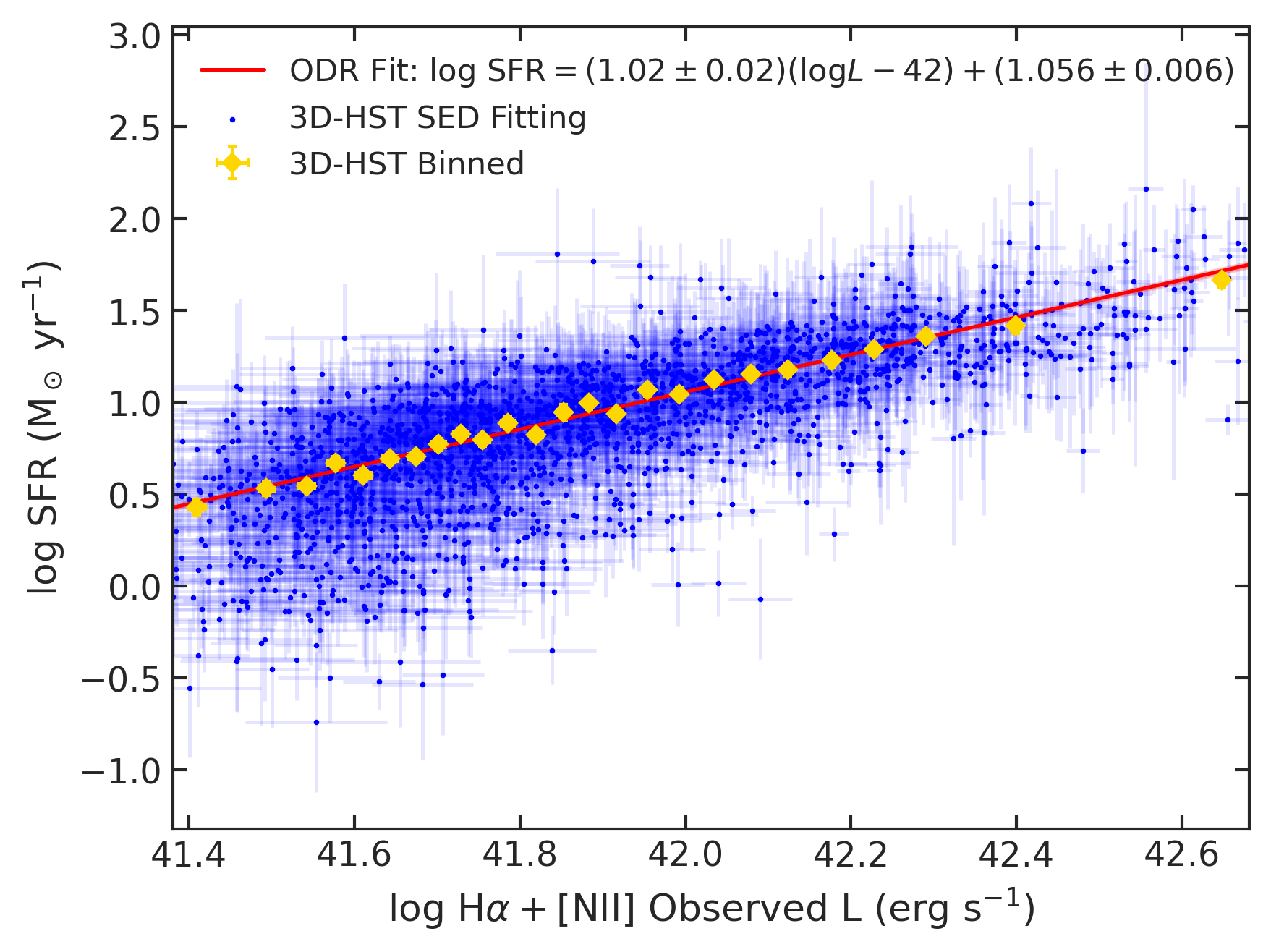}}
    \caption{Calibration between SFR and observed (not corrected for dust) \Ha + \NIIsimp luminosity for the sample used in this paper. We show the data as blue dots, the logarithm of the mean linear SFR and luminosity in 25 bins as amber diamonds, and the best-fit line to the binned mean values in red, which has a slope nearly identical to one (a linear relationship in linear space). There is a clear correlation in the data. }
    \label{fig:sfrhacalib}
\end{figure}

We present the results for our sample in Figure \ref{fig:sfrhacalib}. The data are plotted as blue dots, while the linear mean binned values are shown as amber diamonds. Finally, the best-fit line is shown in red, with the $1\sigma$ uncertainty displayed via the shaded region (nearly too small to be visible). The correlation between apparent \Ha+ \NIIsimp luminosity and SFR is quite clear.


The best-fit relation is given by Equation \ref{eq:sfrhacalib}. The line has a slope consistent with $1$, suggesting that the relation between SFR and uncorrected \Ha+ \NIIsimp luminosity is very close to linear. To be consistent with the SFRD compilation of \cite{MadauDickinson2014}, we have divided the (linear) SFR by $0.67$ as done in their work; this converts an SFR based on the \cite{Kroupa01} IMF to one based on the \cite{Salpeter1955} IMF with limits of 0.1 and $100 \, M_{\odot}$.  In the equation $L_{{\rm H}\alpha+{\rm [N II]}}$ is in units of erg s$^{-1}$ and SFR is in units of $M_\odot$/year.
\begin{equation} \label{eq:sfrhacalib}
    \begin{split}
        \log {\rm SFR} & = (1.02 \pm 0.02) (\log L_{\rm uncorr}-42) \\
        & + (1.056 \pm 0.006) - \log(0.67)
    \end{split}
\end{equation}

To be complete, we also calculate the SFRD using the \cite{Kennicutt2012} relation (shown below in Equation \ref{eq:HaSFR}), after applying corrections for uniform dust attenuation and \NIIsimp contamination.
\begin{equation} \label{eq:HaSFR}
    \log {\rm SFR} = \log L_{{\rm H}\alpha} - 41.27 - \log(0.67)
\end{equation}
Both \cite{Shim2009} and \cite{Colbert2013} use a 29\% correction for \NIIsimp, whereas \cite{Sobral2013} use a formula based on equivalent width and find a median correction of 25\%. The 29\% correction, which we now adopt, is equivalent to a $-0.15$ dex shift in log luminosity. 

\begin{figure}
    \centering
    \resizebox{\hsize}{!}{
    \includegraphics{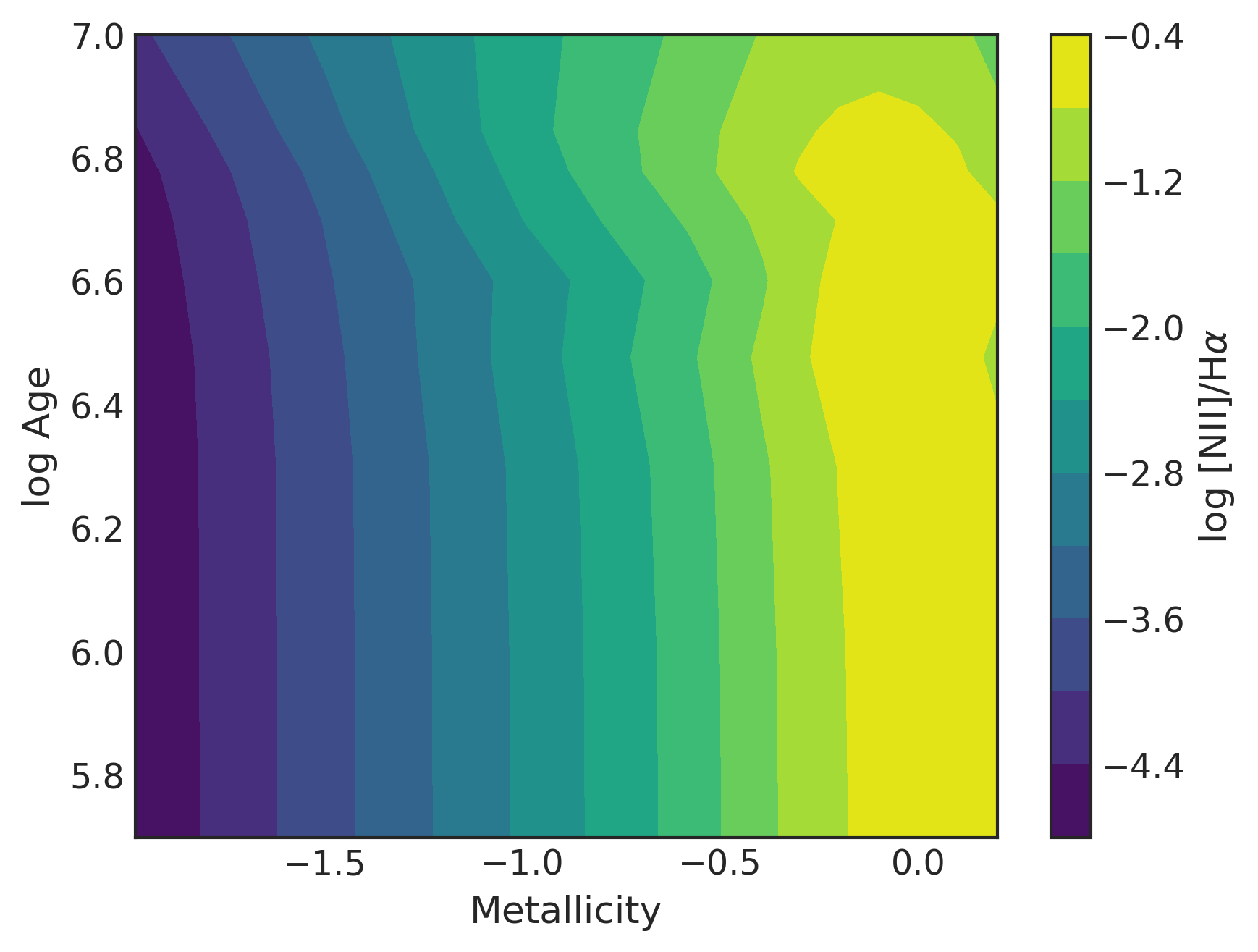}}
    \caption{\cloudy prediction for $\log$ \NIIsimp/\Ha as a function of metallicity and age when $\log U = -2.5$. While $\log$ \NIIsimp/\Ha is nearly independent of age, it is highly dependent on metallicity. For galaxies with $\log(Z/Z_\odot)\lesssim -0.7$, \NIIsimp emission is less than 1\% of the \Ha emission. }
    \label{fig:NIIHaMetAge}
\end{figure}

To test the appropriateness of this 29\% correction ($\log$ \NIIsimp/\Ha $ = -0.54$), we determined how \NIIsimp/\Ha should vary as a function of population age and metallicity using the \cloudy \citep{Ferland1998,Ferland2013} nebular emission tables created by \cite{Byler2017}, under the assumption of $\log U=-2.5$ (the same value used in our \mcsed fits).  We show the result in Figure \ref{fig:NIIHaMetAge}. According to the \cloudy lookup tables, the correction is valid only for most galaxies with $\log(Z/Z_\odot)\gtrsim -0.2$. For galaxies with $\log(Z/Z_\odot)\lesssim -0.2$, the true relative strength of \NIIsimp is lower than the correction.

Stellar metallicity is difficult to measure using SED fits to mostly broadband photometric data \citep[e.g.,][]{Conroy2013,Lower2020}. Nevertheless, from our analysis (see Paper~I) we find that 82\% of the galaxies in our sample have $\log(Z/Z_\odot)< -0.2$; this is consistent with the relatively small stellar masses of the galaxies (median $\log M_*/M_{\odot} \sim 9.5$).  In other words, for the majority of \Ha-emitting galaxies, our 29\% correction overestimates the contamination by \NIIsimp, and thus underestimates the population's SFRD\null. Our \mcsed-calibration bypasses this issue by removing the need to correct for \NIIsimp. 

\cite{Topping2021} do a stacking analysis on $z\sim 1.5$ galaxies found in the Multi-Object Spectrometer for Infra-Red Exploration Deep Evolution Field (MOSDEF) survey. They find that $\log$ \NIIsimp/\Ha is a strong function of stellar mass, which is correlated with metallicity. From their Figure 2, we notice that for galaxies with $\log(M_*/\msun) \lesssim 10.5$, the 29\% correction for \NIIsimp is an overestimate. We find that 81\% of our galaxies have masses $\log(M_*/\msun) < 10.5$, which is in perfect agreement with the aforementioned finding using \cloudy lookup tables.

As for dust, based on results from Paper~I, we find that the average differential extinction, $E(B-V)$, for our galaxy sample is 0.16~mag. To calculate $A({\rm H}\alpha)$ for our sample, we use the conversions given by \cite{Reddy2020},
\begin{align}
    E(B-V)_{\rm nebular} &= 2.07 E(B-V)_{\rm stellar} \\
    A({\rm H}\alpha) &= 2.66 E(B-V)_{\rm nebular}
\end{align}
and apply an average value of $A(H\alpha)=0.88$ to the entire sample. The combination of the \NIIsimp and dust corrections results in a multiplicative factor of $1.60$.

\subsubsection{Star Formation Rate Density Results} \label{subsubsec:sfrdres}

Given the relation between SFR and uncorrected \Ha + \NIIsimp luminosity (Equation \ref{eq:sfrhacalib}), we can calculate the total SFRD using the luminosity function results in this paper. If $\phi(L)$ is the true luminosity function, $\textrm{SFR}(L)$ is the relation given by Equation~\ref{eq:sfrhacalib}, and $L$ is the uncorrected \Ha + \NIIsimp luminosity, then the total SFRD contained in 3D-HST ELGs brighter than some luminosity $L_{\rm min}$ is
\begin{equation} \label{eq:sfrdmcsed}
    {\rm SFRD}(z) = \int_{L_{\rm min}}^\infty {\rm SFR}(L)\, \phi(L,z)\, dL
\end{equation}
Alternatively, we can use the local calibration between SFR and dust-corrected \Ha luminosity (Equation \ref{eq:HaSFR}) and apply the average dust and \NIIsimp corrections as described in \S \ref{subsubsec:conversion}. The net effect of these two factors is included in the constant $k=1.6$, and
\begin{equation} \label{eq:lumtot}
    {\rm SFRD}(z) = {\rm SFR}_{\rm corr}\left(k\, \int_{L_{\rm min}}^\infty L \, \phi(L,z)\, dL \right) 
\end{equation}
\noindent which simplifies to 
\begin{equation} \label{eq:lumtotz}
    {\rm SFRD}(z) = {\rm SFR}_{\rm corr}\left(k\, \Gamma\left(\alpha+2,\frac{L_{\rm min}}{L_*}\right)\, \phi_*(z)\, L_*(z) \right)
\end{equation}
where $\Gamma$ represents the incomplete gamma function.



\begin{figure}
    \centering
    \resizebox{\hsize}{!}{
    \includegraphics{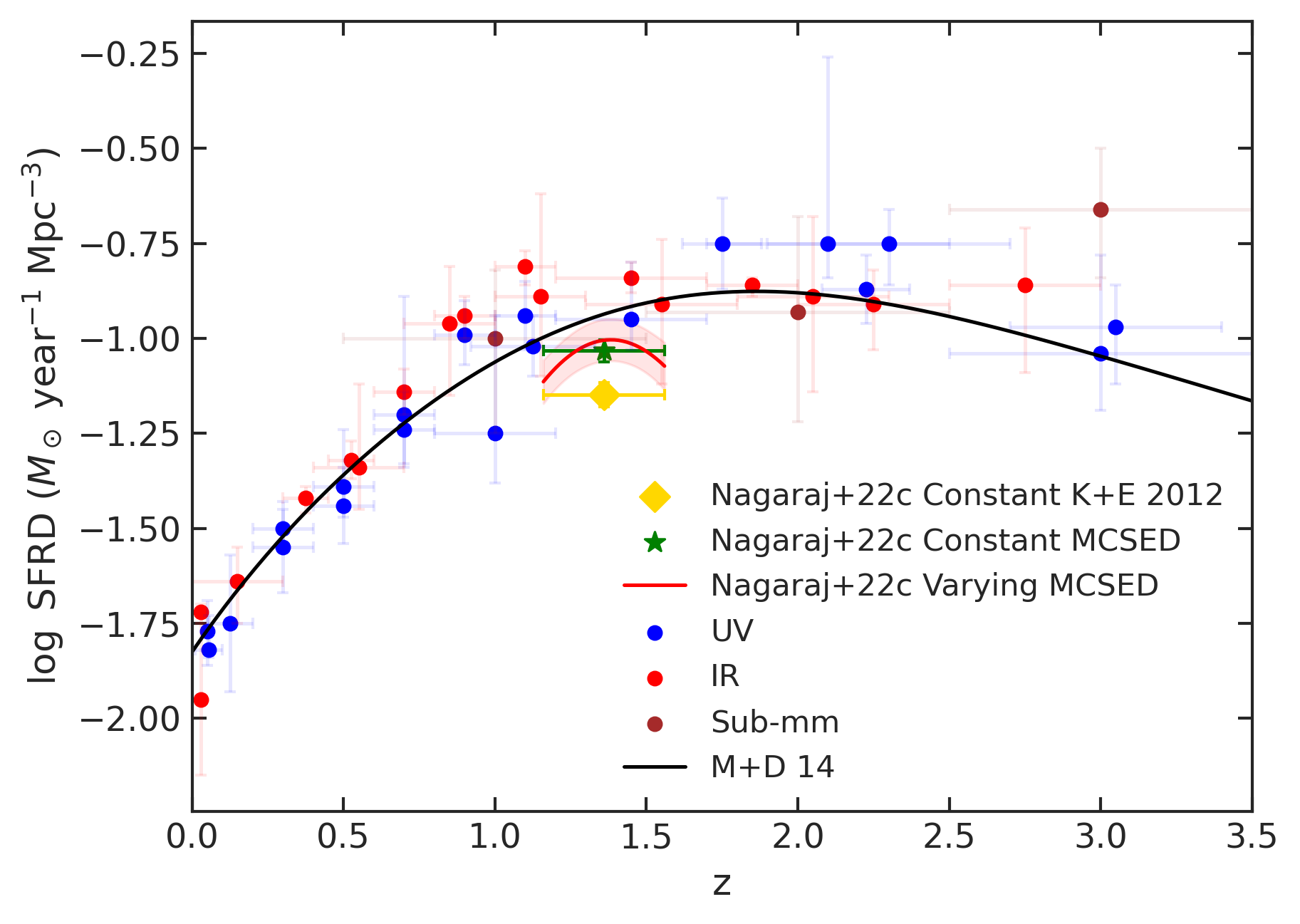}}
    \caption{Our calculated star formation rate density (SFRD) compared to literature values (UV-based results are in blue dots, IR results in red, and sub-mm-based in brown) and the best-fit curve (shown in black) from \cite{MadauDickinson2014}. The green star and red line show the SFRD when using our \mcsed-based \Ha+ \NIIsimp SFR calibration with the static and redshift-varying luminosity functions, respectively. By directly connecting the observed \Ha+\NIIsimp fluxes to SFR, we bypass the issues associated with dust attenuation and \NIIsimp contamination, while propagating in their uncertainties.  The amber diamond shows the SFRD calculated using the non-evolving luminosity function with the \cite{Kennicutt2012} \Ha-SFR calibration; here the uncertainties from dust and \NIIsimp are not included in the error bar.  This latter calibration produces an SFRD that is smaller by $0.12$~dex than SED-based value; this is consistent with the overestimate of \NIIsimp in low-mass (low-metallicity) galaxies. At $z \sim 1.4$, the SFRD from ELGs is $\gtrsim 0.09$ dex lower than the \cite{MadauDickinson2014} curve, but this difference is still within the variance seen in the literature. We are not able to find conclusive evidence of SFRD evolution between $1.16 \leq z \leq 1.56$. }
    \label{fig:SFRDHa}
\end{figure}

Figure \ref{fig:SFRDHa} shows the evolution of the cosmic SFRD based on results compiled by \cite{MadauDickinson2014} from both UV-based and IR-based measurements, along with the best-fit curve from their review paper. We also include SFRD measurements from \cite{Gruppioni2020} using sub-mm data. Our SFRD results for both the non-evolving and redshift-varying luminosity functions, based on our SED-based SFR calibration, are shown as a green star and red line, respectively. Our SFRD measurement derived from the non-evolving luminosity function with the \cite{Kennicutt2012} relation is shown as an amber diamond. In order to make a fair comparison to the literature, we use $L_{\rm min}=0.03\mathcal{L}_*$, as adopted by \cite{MadauDickinson2014}. 

Our measurements derived via the SFR calibration shown in Figure~\ref{fig:sfrhacalib} yield an SFRD that is larger by $0.12$~dex than that found using the local \Ha-SFR calibration given by \cite{Kennicutt2012}. This discrepancy is consistent with the conclusion of Section~\ref{subsubsec:conversion}, i.e., that the application of a 29\% correction for the contribution of \NIIsimp to the \Ha measurement leads to an underestimate in star-formation rate and the epoch's SFRD\null.  We therefore believe that our SED-based SFR calibration is more accurate. However, we do note that the uncertainty on the local-calibration-based SFRD is underestimated since the uncertainties on the dust and \NIIsimp corrections are not propagated into the analysis. Thus, the two numbers may still be consistent. 

We also observe that the cosmic SFRD calculated for our sample of ELGs is at least $\sim 0.09$ dex below the value expected from best-fit curve of \cite{MadauDickinson2014}. This implies that \Ha-selected galaxies contain $\lesssim 81\%$ of the star formation in the $z\sim 1.4$ universe.

The difference between the cosmic SFRD at $z \sim 1.4$, as determined by \cite{MadauDickinson2014}, and our \Ha-based measurement suggests that not all of the epoch's star-formation is detectable via surveys for rest-frame optical emission lines.  Such a result is easily explained if some star-forming galaxies are heavily obscured by dust; since emission-line gas is generally attenuated more than star-light, this is a reasonable hypothesis \citep[e.g.,][]{CharlotFall2000, Calzetti2000, Reddy2020}. These dusty star forming galaxies have been shown to contribute significantly to the star formation rate around cosmic noon, i.e., $1<z<3$ \citep[e.g.,][]{Casey2013}.

At higher redshifts ($z \gtrsim 3$), far-IR and sub-mm studies have shown that UV-based studies miss a significant portion of the star formation due to dust attenuation in extremely high-SFR galaxies \citep[e.g.,][]{Gruppioni2015,Rowan-Robinson2016,Wang2019,Williams2019,Gruppioni2020,Loiacono2020,Khusanova2021}. Nevertheless, the discrepancy between the UV-NIR and FIR-sub-mm methods of calculating star formation \citep[e.g., see the extensive discussion by][]{Katsianis2021} is much less pronounced at the redshifts of our sample. For example, in Figure \ref{fig:SFRDHa}, we include sub-mm results from \cite{Gruppioni2020} and find that for $z=1$ and $z=2$, the results are still consistent with the \cite{MadauDickinson2014} best-fit curve.

Coming back to our results, we cannot discount the possibility that our SFRD is consistent with the \cite{MadauDickinson2014} curve, as the discrepancy is still within the variance of the literature values. Thus, there is no conclusive evidence that the emission line surveys are missing a significant fraction of sources found in UV-based and/or IR/sub-mm-based surveys between redshifts $1.16<z<1.56$.

While the redshift-evolving SFRD is not a flat curve, given the non-negligible uncertainties due to cosmic variance, we cannot dismiss the possibility of no evolution. In other words, our \Ha emission-line measurements find no strong evidence for SFRD evolution over our $1.16<z<1.56$ redshift range.

\subsection{\OIII Luminosity Function} \label{subsec:OIIILF}

Unlike \Ha, \OIII does not suffer from uncertainties due to blending.  At the redshifts under consideration, the G141 grism has enough resolution and the sources are sufficiently small so that \OIIIsimp and \Hb are easily distinguishable.  \OIII is blended with \OIIIl, but since the ratio of the two lines is fixed by basic physics \citep[2.98:1;][]{Storey2000}, converting the observed \OIIIsimp feature into \OIII is trivial.  We quote only the luminosity of \OIII in the subsequent analysis.



\begin{figure}
    \centering
    \resizebox{\hsize}{!}{
    \includegraphics{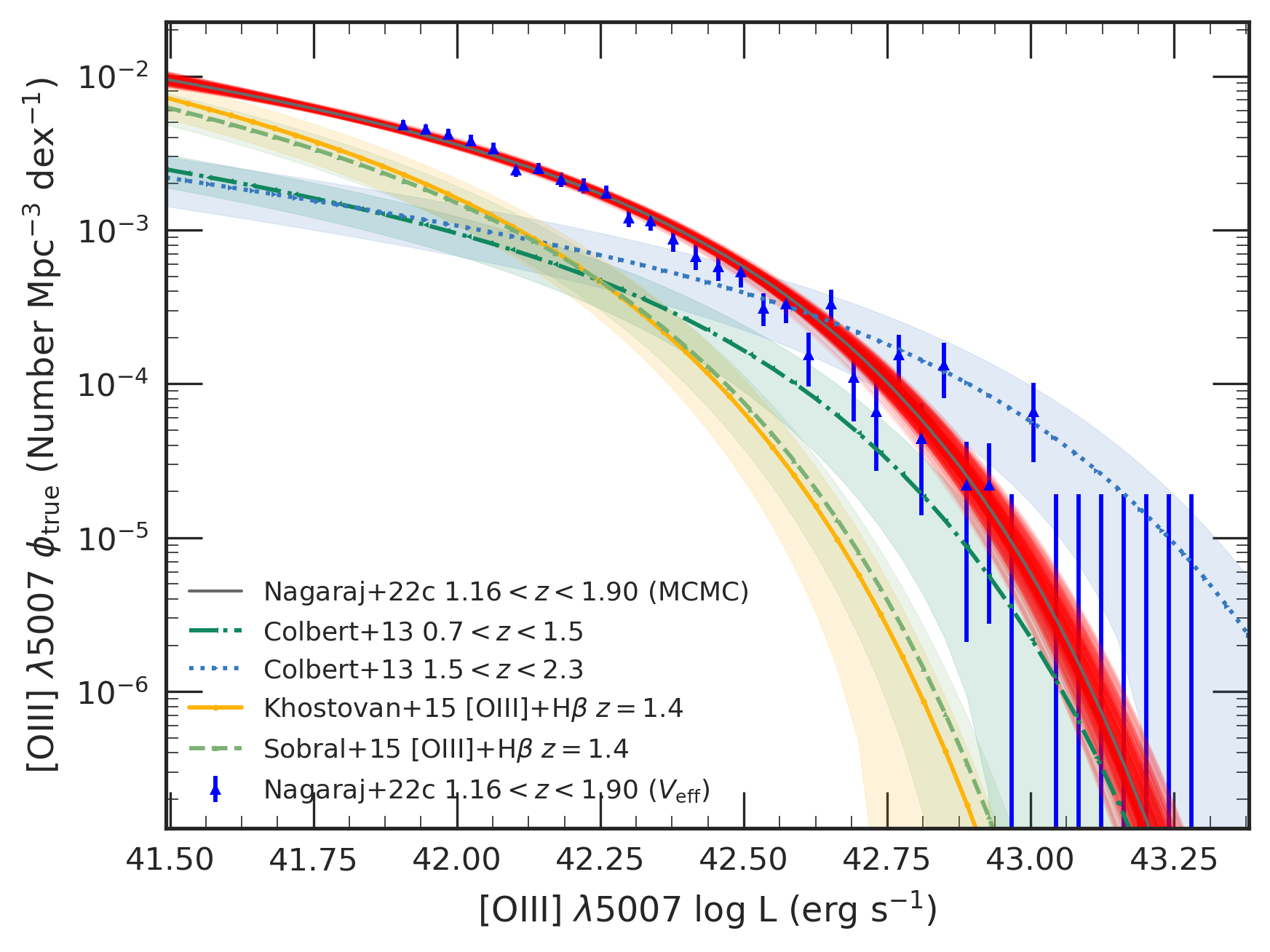}}
    \caption{\OIII luminosity function from this work (MCMC as the red curves and $V_{\rm eff}$ as blue triangles), as well as fits from \cite{Colbert2013}, \cite{Khostovan2015}, and \cite{Sobral2015}. We predict higher counts of intermediate-luminosity galaxies than the other studies, but at the high-luminosity end, our data are well within the range of values previously derived.  There is good agreement between the MCMC and $V_{\rm eff}$ methods. }
    \label{fig:OIIILF}
\end{figure}

In Figure \ref{fig:OIIILF}, we show our fitted non-evolving \OIII luminosity function (over the range $1.16<z<1.90$) and place it in the context of the literature. This includes fits to 192 $0.7<z<1.5$ and 58 $1.5<z<2.3$ \OIIIsimp emitters by \cite{Colbert2013}, 371 $z=1.42$ \OIIIsimp+ \Hb emitters from \cite{Khostovan2015}, and 1343 $z\sim 1.4$ \OIIIsimp+ \Hb emitters from \cite{Sobral2015}. For our comparison, the \OIIIsimp $\lambda \lambda 4959,5007$ values of these works have been converted to \OIII to match our measurements. 

The studies by \cite{Khostovan2015} and \cite{Sobral2015} identify \Hb emitters as well as \OIIIsimp galaxies, since their narrow-band photometry is unable to distinguish the two object classes without spectroscopic follow-up. Thus, it is unclear how their samples compare to ours. \cite{Sobral2015} find that in their set of sources with spectroscopic confirmation, \Hb emitters constitute around 16\% of their $z\sim 1.4$ sample. Interestingly, they notice that these \Hb emitters tend to have lower luminosities than the \OIIIsimp-identified emitters, and this may be reflected in the lower $\mathcal{L}_*$ values implied by the plot. Still 5/6's of their (spectroscopically confirmed) sample is made up of \OIIIsimp emitters, so the effect of these contaminants should not be large. \cite{Khostovan2015} find a similar phenomenon in their sample.

As shown in Figure~\ref{fig:OIIILF}, at $41.5 \lesssim \log L \lesssim 42.3$, our luminosity function predicts more \OIIIsimp galaxies than any of the other studies, though at higher luminosities our results are well within the bounds of the literature. Moreover, our MCMC result is in good agreement with the $V_{\rm eff}$ result (blue triangles). 

There are a few factors that may lead to our distinct result for \OIIIsimp. The first is that our redshift range is unique. No other study focuses specifically on $1.16<z<1.90$ galaxies. As the luminosity function is an evolving quantity, the particular redshifts involved play a role in determining the measured parameters. Another important consideration is that our sample size is the largest to date, and the larger the sample size, the more accurate the luminosity function. In fact, as shown in Table \ref{tab:OIIILumFunc}, the value for $\log\, \int_{0.03L^*}^\infty \phi(L)\, dL$ derived by \cite{Sobral2015} is closest to our study, and it is also the measurement that is most consistent with our value.

Alternatively, we note that our analysis encounters problems at both the faint and bright ends of the luminosity function. At the faint end, we are subject to rapid loss of completeness while at the bright end, our measurements of the luminosity function suffer from cosmic variance as we are dealing with small numbers of galaxies. Coupled with degeneracies between Schechter parameters, these effects can lead to divergences between different studies. Extrapolations to low-luminosity galaxies, especially when $\alpha$ is fixed, are often uncertain and should be considered as such.  Nevertheless, given our large sample size and the number of galaxy measurements at luminosities significantly less than $L_*$, we believe that the differences between our results and those of previous studies are real and not an artifact of our analysis.


\begin{figure*}
    \centering
    \resizebox{\hsize}{!}{
    \includegraphics{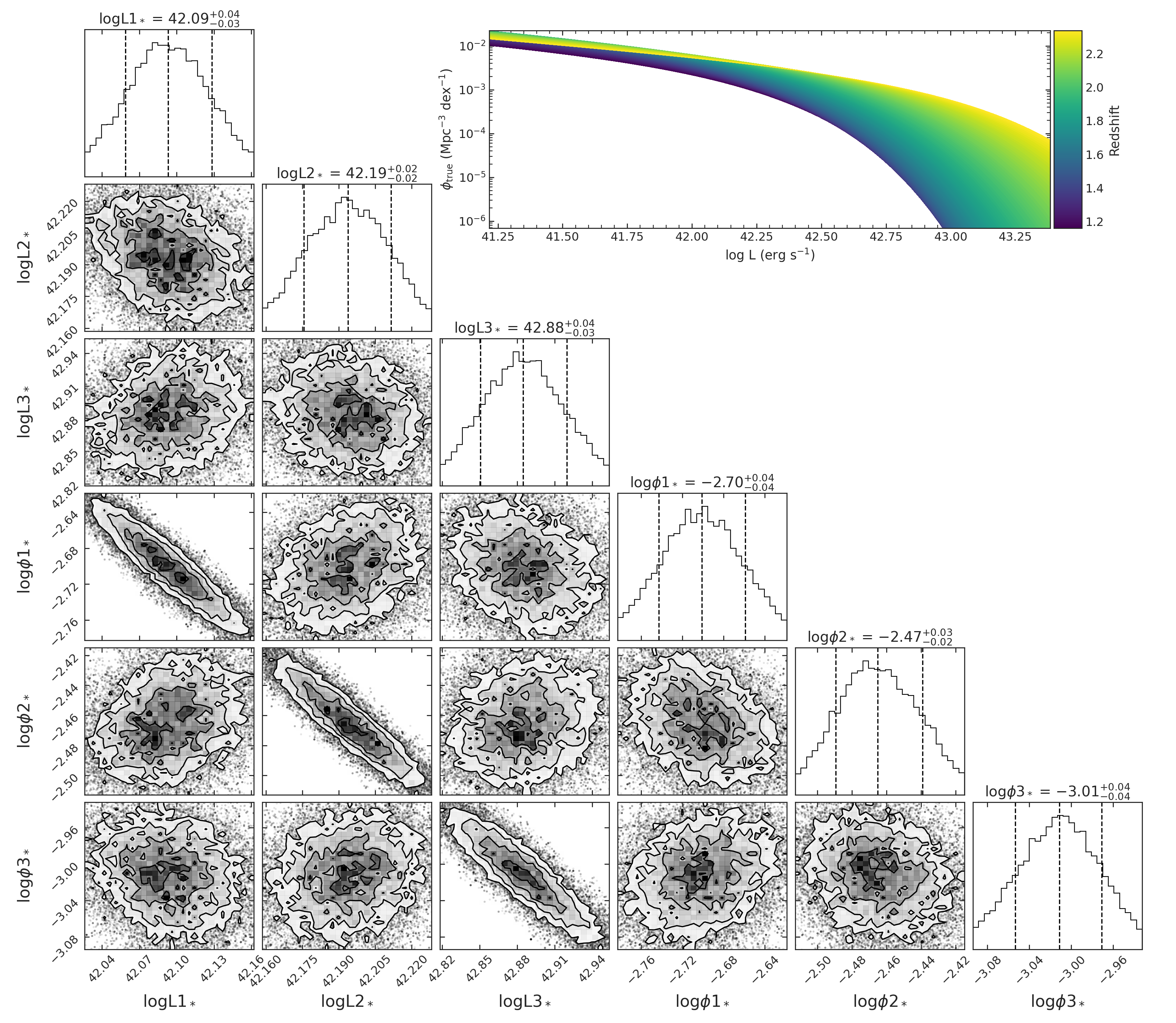}}
    \caption{MCMC result for the redshift varying \OIII luminosity function. In the lower left, we show the 1D and 2D cross sections of the parameter chains.  Note that at each redshift, $\mathcal{L}_*$ and $\log \phi_*$ are correlated, but the correlation is not strong across redshifts.  The upper-right panel shows the evolution of the luminosity function.  Our data show that $\mathcal{L}_*$ increases with redshift, and, at most luminosities, the luminosity function is larger at higher redshift. In other words, \OIII-bright galaxies were much more common at $z\sim 2$ than at $z\sim 1$. }
    \label{fig:OIIIzLF}
\end{figure*}

In Figure \ref{fig:OIIIzLF}, we combine our \OIII measurements with those of \cite{Bowman2021} to show the redshift evolution of the \OIII luminosity function between $1.16\leq z \leq 2.35$.  For this analysis, we fix $\alpha=-1.5$, as it very difficult to fit $\alpha$ as a free parameter when the limiting luminosity depends so much on redshift. Like in Figure \ref{fig:HaTriangle}, we show the 1D and 2D cross sections of the MCMC chains in the lower left panels.  Once again, we find that $\mathcal{L}_*$ and $\log \phi_*$ are correlated, but only at the same redshift. In other words, $\mathcal{L}_{1*}$ and $\log \phi_{1*}$ are highly correlated but $\mathcal{L}_{1*}$ and $\log \phi_{2*}$ are not. 

The data of Figure~\ref{fig:OIIIzLF} show that the characteristic luminosity $\mathcal{L}_*$ increases with redshift. Moreover, as redshift increases, the overall luminosity function increases at all but the lowest luminosities. In other words, there are many more \OIII-visible galaxies, and especially more \OIII-bright systems, at earlier epochs of cosmic history.  This is consistent with the findings of \cite{Zeimann2014}, \cite{Khostovan2015},  and \citet{Bowman2019}, among others, that show the prevalence of \OIIIsimp compared to \OII at high redshift.

\subsection{Number Counts} \label{subsec:numcounts}

The precision to which one can measure cosmological parameters through galaxy surveys depends on the number of galaxies and the square of the bias of the observed galaxy population relative to dark matter. Surveys planned for \textit{Euclid} and \textit{Roman} will observe millions of \Ha-visible galaxies at $0.9 \lesssim z \lesssim 1.8$ and \OIIIsimp-visible galaxies at $1.5 \lesssim z \lesssim 2.7$, and thus measure quantities such as the angular diameter distance and Hubble parameter at these distant epochs.  Specifically, the \textit{Euclid} Wide Survey (WS) will observe $\sim 15000~{\rm deg}^2$ of the sky down to a flux limit of $\sim 2 \times 10^{-16}$ \flux \citep{Scaramella2021}, while the \textit{Roman} High Latitude Survey (HLS) will observe $\sim 2200~{\rm deg}^2$ down to a flux limit of $\sim 6 \times 10^{-17}$ \flux \citep{Spergel2015}. The \textit{Euclid} Deep Survey will observe $50~{\rm deg}^2$ in three fields to a similar flux limit \citep{Vavrek2016}.  

In this section, we use our measurements of the \Ha and \OIII luminosity functions to calculate the number of galaxies these surveys are likely to measure, corrected for completeness.  One factor to note is that since we have removed AGN from the sample, the number counts we derive will be a slight underestimate.  

We begin with the total number of galaxies with luminosities greater than $L_{\rm min}(z)$ from redshifts $z_0$ to $z_1$ over area $\Omega$,
\begin{equation} \label{eq:numtot}
    N_{\rm tot} = \int d\Omega \int_{z_0}^{z_1} dz \frac{dV}{dz \, d\Omega} \int_{L_{\rm min}(z)}^\infty dL \, \Omega(L,z) \phi(L,z)
\end{equation}
Given the Schechter function with parameters $\alpha$, $L_*$, and $\phi_*$ (that are assumed to be invariant over the effective survey area $\Omega_0$), we can simplify Equation \ref{eq:numtot} to
\begin{equation} \label{eq:numtotz}
    N_{\rm tot} = \Omega_0 \int_{z_0}^{z_1} dz \frac{dV}{dz d\Omega} \, \Gamma\left(\alpha+1,\frac{L_{\rm min}(z)}{L_*}\right) \phi_*(z) L_*(z)
\end{equation}
where $\Gamma$ once again represents the incomplete gamma function.  

\begin{figure*}
    \centering
    \resizebox{\hsize}{!}{
    \includegraphics{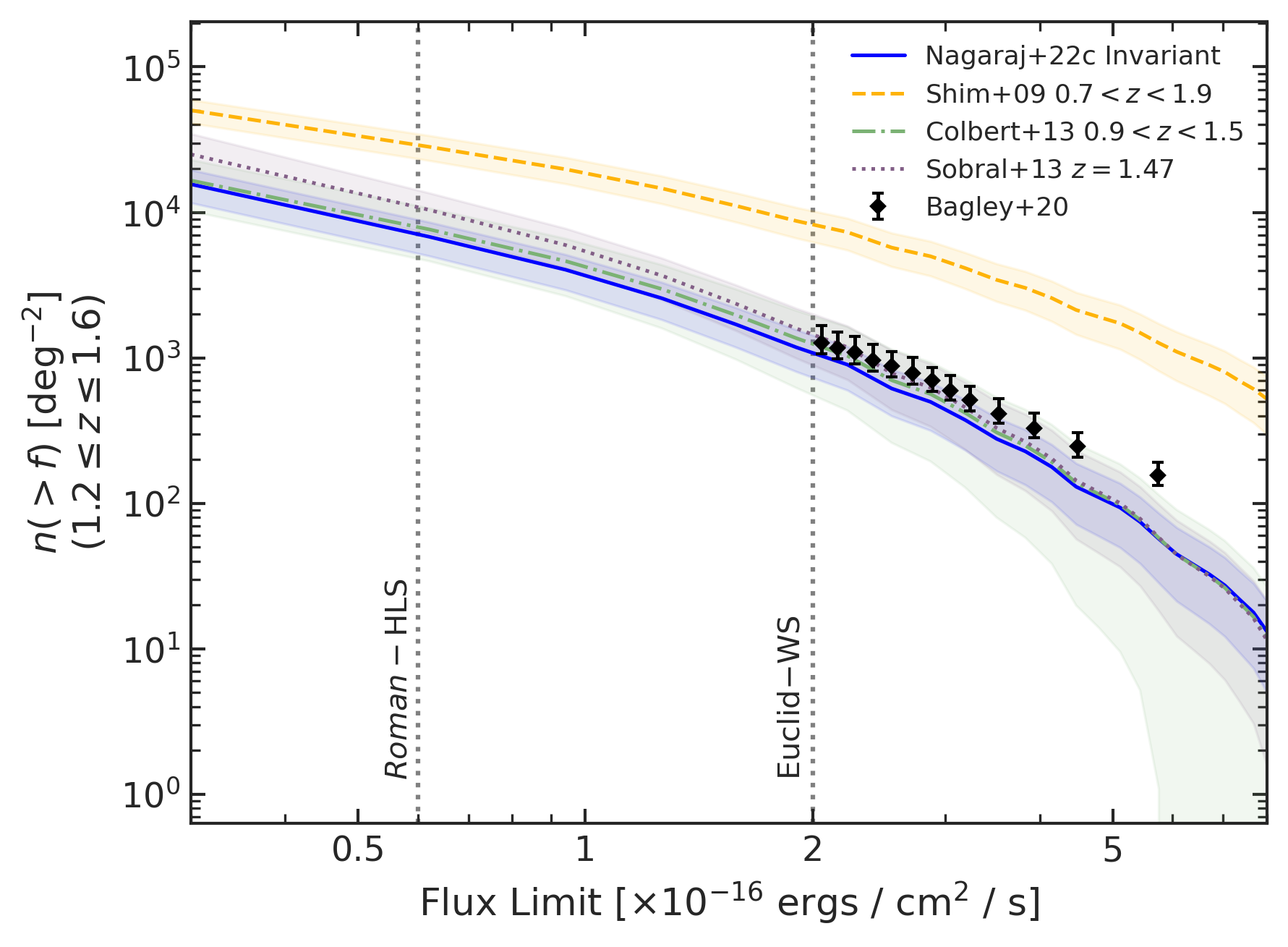}
    \includegraphics{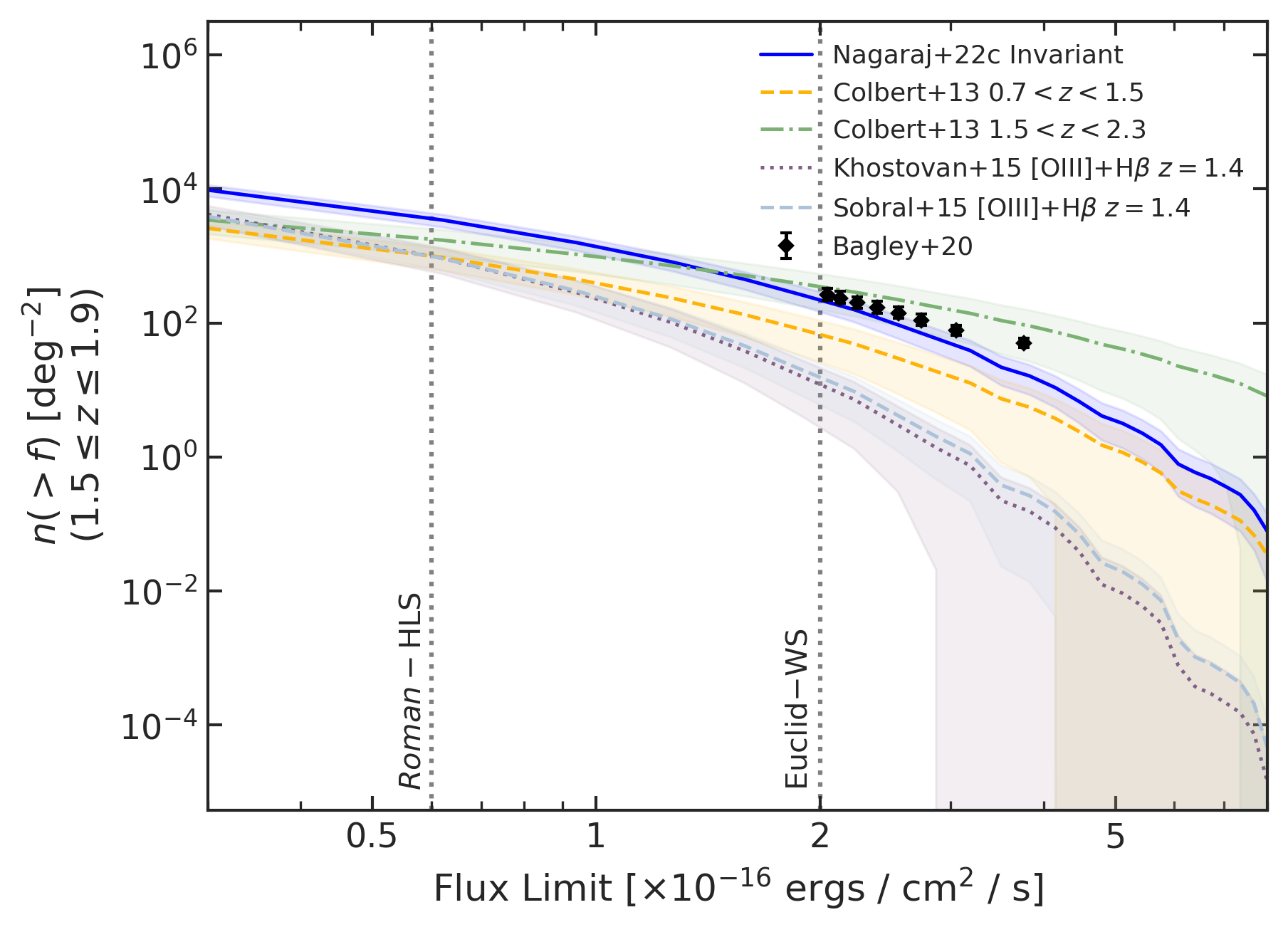}}
    \caption{Predictions for total number of galaxies per square degree corrected for completeness, as a function of the limiting flux for \Ha at $1.2\leq z \leq 1.6$ (left) and \OIII at $1.5\leq z \leq 1.9$ (right). We use our non-evolving luminosity functions for these calculations and include cosmic variance in the error budget. Our \Ha counts agree remarkably well with those from \cite{Colbert2013} at all limiting fluxes as well as with the direct counts from \cite{Bagley2020} around $F_{\rm lim}\sim 2 \times 10^{-16}$ \flux. This agreement suggests no changes to the \Ha predictions for \textit{Euclid} and \textit{Roman} are necessary. While our \OIII count predictions are distinct from the literature, they still do not greatly change the expected counts from \textit{Euclid}. However, for the HLS flux limit of $6.0 \times 10^{-17}$ \flux, we predict twice as many galaxies as \cite{Colbert2013}. }
    \label{fig:IntegCounts}
\end{figure*}

In Figure \ref{fig:IntegCounts}, we show the total number of galaxies per square degree that are expected to have \Ha at $1.2\leq z \leq 1.6$ (left) and \OIII at $1.5\leq z \leq 1.9$ (right) above a given threshold.  We calculate these values using Equation \ref{eq:numtotz} with $\Omega_0=1$ deg$^2$, while translating $L_{\rm min}$ to emission line flux using the luminosity distance $D_L$ via
\begin{equation}
    F_{\rm min} = \frac{L_{\rm min}(z)}{4\pi D_L(z)^2}
\end{equation}
For the figure, we use the non-evolving luminosity functions (though the redshift varying luminosity functions yield similar results) and perform the same calculations for the luminosity functions given in the literature.  In all cases, the values represent galaxy number counts, assuming 100\% completeness. In addition, we include the direct measurements of counts obtained by \cite{Bagley2020} using samples of ELGs from the WFC3 Infrared Parallel Spectroscopic Survey (WISPS), 3D-HST, and A Grism H-Alpha SpecTroscopic survey (AGHAST)\null. This work also made corrections for completeness.

In the left panel, we observe that our \Ha counts are in excellent agreement with those of \cite{Colbert2013} at all values of the limiting flux. The counts also agree with those of \cite{Bagley2020} around the limiting flux of the \textit{Euclid} WS, but less so at brighter fluxes. Given the consistency of our results with those of \cite{Colbert2013} and \cite{Bagley2020}, there is no need to update the prediction of $\sim 3300$ deg$^{-2}$ for the \textit{Euclid} WS at 100\% completeness \citep{Bagley2020,Scaramella2021} and 16.4 million $7\sigma$ \Ha galaxy detections at $1.06<z<1.88$ for the \textit{Roman} HLS assuming 70\% completeness \citep{Spergel2015}.

On the other hand, our result for \OIIIsimp (right side of Figure \ref{fig:IntegCounts}) is distinct from the literature. For limiting fluxes of 1 to $3 \times 10^{-16}$ \flux, our predicted galaxy counts agree with the results of \cite{Colbert2013} and \cite{Bagley2020}. As these limits are similar to those of the \textit{Euclid} WS ($2 \times 10^{-16}$ \flux) and the $\geq 7\sigma$ limit of $10^{-16}$ \flux used by \cite{Spergel2015} for predicting \textit{Roman} HLS counts, our results do not change the predictions for these surveys; at 70\% completeness, the \textit{Roman} HLS should detect 1.4 million \OIIIsimp $1.88<z<2.77$ galaxies with $> 7\sigma$ confidence \citep{Spergel2015}. However, as we push to lower flux limits, our luminosity function predicts higher counts. For example, for the \textit{Roman} HLS nominal flux limit of $6.0 \times 10^{-17}$ \flux, we predict twice as many galaxies as \cite{Colbert2013} in the redshift range $1.5<z<1.9$. 

\section{Conclusion}\label{sec:disc}

In Papers~I and II, we used the G141 grism data from the \textit{Hubble} 3D-HST Treasury program \citep[GO-11600, 12177, 12328;][]{Brammer2012,Momcheva2016} to identify a clean sample of 4350 normal (i.e., non-AGN) star-forming  emission lines galaxies in the redshift range $1.16 \leq z \leq 1.90$.  In this work, we use the line fluxes of these galaxies, along with a similar set of fluxes measured by \citet{Bowman2021}, to measure the galaxies' emission-line luminosity functions. These data include 1892 \Ha emitting galaxies between $1.16\leq z \leq 1.56$ and 4519 \OIIIsimp emitters with $1.16 \leq z \leq 2.35$, all with line fluxes above the 3D-HST 50\% completeness limit.  While there have been several previous efforts to calculate \Ha and \OIIIsimp luminosity functions in these redshift ranges (see \S \ref{sec:intro} for a detailed list), none used the large sample sizes used here. 


We employ a generalization of the classical $1/V_{\rm max}$ method to derive the emission-line luminosity functions for our entire sample of galaxies, and samples of galaxies broken down by redshift.  We then use Markov Chain Monte Carlo (MCMC) Bayesian techniques to fit these data to the \cite{Schechter1976} luminosity function with $\alpha$ held constant across redshift.   We find very good agreement between our \Ha results and those from the literature, and our \OIIIsimp luminosity function is also a good match to prior measurements for line luminosities brighter than $\log L = 42.3$ (ergs~s$^{-1}$).  However, at fainter \OIIIsimp luminosities (where completeness corrections might be an issue), we infer an excess of objects.  These results are shown in Figures \ref{fig:HaTriangle}, \ref{fig:HazLF}, \ref{fig:OIIILF}, and \ref{fig:OIIIzLF} and summarized in Tables \ref{tab:HaLumFunc} - \ref{tab:EvolvingBest}.

We also compute the star formation rate density (SFRD) of the $1.16 \leq z \leq 1.56$ epoch using the \Ha luminosity function.  We find that our SFRD is $\sim19\%$ smaller than the best-fit $z\sim 1.4$ value found by \cite{MadauDickinson2014} (Figure \ref{fig:SFRDHa}), though this discrepancy is within the variance found in the literature.  If the difference is real, then one possible explanation is that not all $z \sim 1.4$ star formation takes place in galaxies with observable \Ha emission lines. In particular, surveys such as 3D-HST will miss heavily obscured galaxies where the emission lines are too extinguished to make it into the sample. We find no evidence for or against cosmic evolution of the SFRD between $1.16<z<1.56$



Finally, we predict total galaxy counts per square degree as a function of the limiting flux (Figure \ref{fig:IntegCounts}). For \Ha + \NII, our results are consistent with those from \cite{Colbert2013} at all limiting fluxes and \cite{Bagley2020} down to a limiting flux of $\sim 2 \times 10^{-16}$ \flux, suggesting the previous predictions for the \textit{Euclid} \citep{Scaramella2021} and \textit{Roman} \citep{Spergel2015} surveys are accurate.  For \OIII, our numbers agree with previous estimates for the \textit{Euclid} Wide Survey, but depending on where exactly we define the flux limit for the \textit{Roman} High Latitude Survey, our data may imply a significantly larger number of detectable galaxies.  For example, at $10^{-16}$ \flux, our results are consistent with the analysis of \cite{Colbert2013}, but at $6 \times 10^{-17}$ \flux, our number counts are higher by a factor of two.  \textit{Roman} may find more faint \OIIIsimp emitters than previously anticipated.

The $\Lambda$CDM paradigm has garnered many resounding successes in explaining observations of our universe at a variety of scales. However, there are still inconsistencies and unknowns leaving the cosmological model incomplete. To better constrain the model as well as alternate or additional theories, we need to continue honing our observations. Large galaxy surveys represent an important avenue to constrain cosmological parameters through the measurement of baryonic acoustic oscillations and redshift space distortions.

Galaxy surveys with precise redshifts will be especially useful for generating the necessary constraints, and IFU and slitless spectroscopy are the most efficient ways of performing these surveys. In the near future, \textit{Euclid} and \textit{Roman} will greatly enhance samples of emission line galaxies. The similarities between 3D-HST and these planned surveys make it a perfect pathfinder mission. Our measurements of the \Ha and \OIII luminosity functions with galaxy samples that are several times larger than any previous study of the $z \sim 1.5$ redshift range help cement the predictions for the expected yield of the ongoing and future surveys.

\acknowledgments

We thank the anonymous referee for their insightful advice that helped make the paper more thorough. This work has made use of the Rainbow Cosmological Surveys Database, which is operated by the Centro de Astrobiolog{\'i}a (CAB/INTA), partnered with the University of California Observatories at Santa Cruz (UCO/Lick,UCSC). This work is based on observations taken by the CANDELS Multi-Cycle Treasury Program with the NASA/ESA HST, which is operated by the Association of Universities for Research in Astronomy, Inc., under NASA contract NAS5-26555. This research has made use of NASA’s Astrophysics Data System. This research has made use of the SVO Filter Profile Service (http://svo2.cab.inta-csic.es/theory/fps/) supported from the Spanish MINECO through grant AYA2017-84089. 

Computations for this research were performed on the Pennsylvania State University’s Institute for Computational and Data Sciences’ Roar supercomputer. The Institute for Gravitation and the Cosmos is supported by the Eberly College of Science and the Office of the Senior Vice President for Research at the Pennsylvania State University.

This material is based upon work supported by the National Science Foundation Graduate Research Fellowship under Grant No.~DGE1255832. Any opinion, findings, and conclusions or recommendations expressed in this material are those of the authors(s) and do not necessarily reflect the views of the National Science Foundation.


%

\vspace{5mm}
\facilities{HST (WFC3), Spitzer (MIPS), Herschel (PACS, SPIRE), GALEX, Swift(UVOT)}


\software{AstroPy \citep{Astropy2013,Astropy2018}, SciPy \citep{Scipy2001,Scipy2020}, CLOUDY \citep{Ferland1998,Ferland2013}, FSPS \citep{Conroy2009,Conroy2010SPSM}, MCSED\citep{Bowman2020} }



\clearpage

\bibliography{sample63}{}
\bibliographystyle{aasjournal_mod}



\end{document}